\documentclass[nofootinbib,aps,preprint]{revtex4}
\usepackage{axodraw}
\usepackage{amsmath}
\usepackage{amssymb}
\usepackage[dvips]{graphics,color}
\newcommand{\nbb}[0]{{0\nu}\beta\beta}
\newcommand{\lbb}[0]{\Lambda_{\beta\beta}}
\newcommand{\lh}[0]{\Lambda_{\text{H}}}
\newcommand{\gf}[0]{G_{\text{F}}}

\newcommand{\Eq}[1]{Eq.~(\ref{#1})}

\newcommand{\Fig}[1]{Fig.~\ref{#1}}

\def\notder{\not\!\partial}

\begin{document}


\vspace{1cm}

\title{Neutrinoless Double $\beta$-Decay and Effective Field Theory}

\author{G. Pr\'{e}zeau$^{a}$, M. Ramsey-Musolf$^{a,b,c}$, and Petr Vogel$^{a}$
\\
}

\vspace{0.3cm}

\affiliation{\scriptsize
$^a$ Kellogg Radiation Laboratory,
California Institute of Technology, Pasadena, CA 91125\ USA\\
$^b$ Department of Physics,\nolinebreak[4]University of
Connecticut,\nolinebreak[4]
Storrs, CT
06269, USA\\
$^c$ Institute of Nuclear Theory, University of Washington,
 Physics/Astronomy Building Box 351550,
 Seattle, WA 98195-1550\ USA\\
}


\begin{abstract}

We analyze neutrinoless double $\beta$-decay ($\nbb$-decay)
 mediated by heavy particles  from the standpoint of
effective field theory.  We show how symmetries of the $\nbb$-decay
 quark operators arising in a given particle physics model determine
 the form of the corresponding effective, hadronic operators.
We classify the latter according to their symmetry transformation
 properties as well as the order at which they appear in a derivative
 expansion.  We apply this framework to several particle physics
 models, including R-parity violating supersymmetry (RPV SUSY) and the
 left-right symmetric model (LRSM) with mixing and a right-handed
 Majorana neutrino.  We show  that, in general, the pion exchange
 contributions to $\nbb$-decay dominate over the
short-range four-nucleon operators.  This confirms previously
 published RPV SUSY results and allows us to derive  new constraints
 on the masses in the LRSM.  In particular, we
show how a non-zero mixing angle $\zeta$ in the left-right symmetry model
produces a new potentially dominant contribution to $\nbb$-decay that
substantially modifies  previous
limits on the masses of the right-handed neutrino and boson stemming
from constraints from $\nbb$-decay and vacuum stability requirements.

\end{abstract}


\maketitle

\pagenumbering{arabic}

\section{Introduction}\label{sec1}

\suppressfloats[t]


\begin{center}
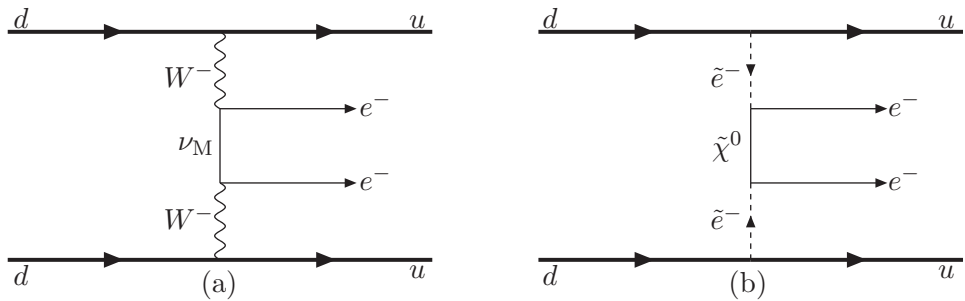
\begin{figure}\caption{ \sl \footnotesize a) $\nbb$ through the exchange of a
  Majorana neutrino.  b)
  $\nbb$ through the exchange of two selectrons and a neutralino
  in RPV SUSY.}\label{intrographs}
\begin{picture}(360,120)(0,0)
%
%
{\SetWidth{1.5}
\ArrowLine(0,90)(80,90)
\ArrowLine(80,90)(160,90)
\ArrowLine(0,4)(80,4)
\ArrowLine(80,4)(160,4)}
\Photon(80,90)(80,61){2}{4}
\Photon(80,4)(80,33){2}{4}
\LongArrow(80,61)(130,61)
\LongArrow(80,33)(130,33)
\Line(80,61)(80,33)

\Text(5,92)[b]{$d$}
\Text(155,92)[b]{$u$}
\Text(5,2)[t]{$d$}
\Text(155,2)[t]{$u$}
\Text(78,74)[r]{$W^-$}
\Text(78,20)[r]{$W^-$}
\Text(133,63)[l]{$e^-$}
\Text(133,35)[l]{$e^-$}
\Text(80,0)[t]{(a)}
\Text(78,47)[r]{$\nu_{\text{M}}$}

%
%

{\SetWidth{1.5}
\ArrowLine(200,90)(280,90)
\ArrowLine(280,90)(360,90)
\ArrowLine(200,4)(280,4)
\ArrowLine(280,4)(360,4)}
\DashArrowLine(280,90)(280,61){2}
\DashArrowLine(280,4)(280,33){2}
\LongArrow(280,61)(330,61)
\LongArrow(280,33)(330,33)
\Line(280,61)(280,33)

\Text(205,92)[b]{$d$}
\Text(355,92)[b]{$u$}
\Text(205,2)[t]{$d$}
\Text(355,2)[t]{$u$}
\Text(278,74)[r]{${\tilde{e}}^-$}
\Text(278,20)[r]{${\tilde{e}}^-$}
\Text(333,63)[l]{$e^-$}
\Text(333,35)[l]{$e^-$}
\Text(278,47)[r]{${\tilde{\chi}}^0$}
\Text(280,0)[t]{(b)}
\end{picture}
\end{figure}
\end{center}

The study of neutrinoless double beta-decay
($\nbb$-decay) is
an important topic in particle and nuclear physics (for recent
reviews, see Refs.
\cite{Elliott:2002xe,Vergados:pv,Vogel:2000vc}).  The
discovery of neutrino oscillations in atmospheric, solar and reactor
neutrino experiments proves the existence of a non-vanishing neutrino
mass~\cite{Fukuda:1998mi,Ahmad:2001an,Eguchi:2002dm}. While
oscillation experiments provide
information on mass-squared differences, they cannot by themselves
determine the
magnitude of the neutrino masses nor determine if neutrinos are
Majorana particles. If the neutrino sector of an \lq\lq extended"
Standard Model includes massive, Majorana
neutrinos, then  $\nbb$-decay provides direct information on the
Majorana masses.  Indeed, since Majorana neutrinos violate lepton number
($L$), Feynman graphs such as the one depicted in
\Fig{intrographs}a are
non-vanishing.  In particular, if the $e,\mu,\tau$ neutrinos have
non-vanishing Majorana masses, an analysis of $\nbb$ coupled with data
from neutrino oscillations provides limits on the absolute value of
these light neutrino masses~\cite{Pascoli:2002xq}.

Neutrinoless $\beta\beta$-decay can also be a probe for heavy mass
scales.  For example, in the
left-right symmetric
model~\cite{Vergados:pv,Mohapatra:1974gc,Senjanovic:1975rk},
a heavy right-handed neutrino
also contributes to the process; it can even be dominant depending on
the values of the elements of the mixing matrix.  Thus, $\nbb$ can be
a tool for the exploration of energy scales beyond the electroweak
symmetry breaking scale.  Alternatively, the
$L$-violating interactions responsible for $\nbb$-decay may not involve
Majorana neutrinos directly.
For example, semileptonic, R parity-violating (RPV) supersymmetric (SUSY)
interactions, involving
exchange of charged-lepton superpartners (an example of which is given
in \Fig{intrographs}b rather than Majorana neutrinos,
can give rise
to~$\nbb$-decay~\cite{Mohapatra:su,Vergados:1986td,Hirsch:1995ek}.
Here again  $\nbb$-decay provides a
probe of the heavy SUSY mass scale and imposes
constraints on RPV SUSY parameters \cite{Hirsch:zi}.  Furthermore, these
alternative scenarios for $\nbb$-decay are relevant for the
study of Majorana neutrinos since any  $\nbb$-decay mechanism will
generate Majorana masses for the neutrinos  \cite{Schechter:1981bd}.

The left-right symmetric model and RPV SUSY are but two of a number of
models that involve a heavy mass scale $\lbb$ that characterizes
the heavy, $L$-violating physics.
Although the effects of these mechanisms will typically be suppressed
by some inverse
power of $\lbb$, $\nbb$-decay mediated by light neutrinos can also be
suppressed since the amplitude is proportional to the neutrino
effective mass.  Thus, it is
important to analyze systematically the potentially comparable
contributions stemming from $L$-violating mechanisms mediated by
heavy  particles.  Since $\lbb$ is far heavier than any hadronic
scale that would enter the problem, there exists a clear separation of
scales in this case.  For the analysis of such situations, effective
field theory (EFT) is the tool of choice.

In what follows, we systematically organize the $\nbb$-decay problem
using EFT, focusing  on
$L$-violation  mediated by heavy physics~(for other efforts along
these lines, see Refs.~\cite{Pas:1997cp,Pas:fc,Pas:2000vn}). Since the
particle physics  dynamics of this heavy
physics occur primarily at
short-distance, one may \lq\lq integrate out" the heavy degrees of freedom,
leaving an effective theory
of quarks and leptons; these quark-lepton operators in turn generate
hadron-lepton operators that have the same transformation properties under
various symmetries.  In this work, only the lightest quarks are
considered, with the relevant symmetries being parity and strong
SU(2$)_{\text{L}}\times$SU(2$)_{\text{R}}$ [chiral SU(2)].  The
effective hadron-lepton
Lagrangian for this theory, ${\cal{L}}_{\text{EFF}}^{\nbb}$, contains
an infinite tower of non-renormalizable operators, which may be
systematically classified in powers of
$p/\Lambda_{\text{H}}$, $p/{\lbb}$ and
$\Lambda_{\text{H}}/{\lbb}$.  Here, $p$ denotes any
small quantity, such as $m_\pi$ or the energy of the
dilepton pair and
$\Lambda_{\text{H}}\sim 1$ GeV is a hadronic mass scale. While the
coefficients of the effective operators in
${\cal{L}}_{\text{EFF}}^{\nbb}$ are unknown\footnote{The computation
  of these coefficients from the underlying quark-lepton interaction
  introduce some degree of uncertainty -- a problem we will not address
  in this work.}, the symmetry
properties of the underlying short-distance
physics may require that certain operator coefficients vanish.

These symmetry properties can have significant consequences for the size of
$\nbb$-decay nuclear matrix
elements and, thus, for the short-distance mass scale deduced from
experimental limits.
Specifically, the hadronic vertices appearing in
${\cal{L}}_{\text{EFF}}^{\nbb}$ will be of the type $NNNNee$,
$NN\pi ee$ and $\pi\pi ee$, etc.    They stem from
quark-lepton operators having different transformation properties
under parity and chiral SU(2); as such,
they will contribute to different orders in the
$p/\Lambda_{\text{H}}$ expansion.

Traditionally, the short-range $NNNNee$ contribution to $\nbb$-decay has
been  analyzed using a
form-factor approach~\cite{vergados81c} where the finite size of
the nucleon is taken into account with the use of a dipole
form-factor.  The form-factor overcomes the
short-range repulsive core in $NN$ interactions that would otherwise
prevent the nucleons from ever getting close enough to exchange the
heavy particles that mediate $\nbb$-decay.  The disadvantage of a
form-factor model is that the error introduced by
the  modeling cannot be estimated systematically in contrast to the
EFT approach.  A discussion of the $NNNNee$ vertex within the
framework of EFT will appear later in this paper.

In contrast to the short range contribution to
$\nbb$-decay, the long range contributions involve the exchange of 
pions~\cite{Pontecorvo:wp} through the $NN\pi ee$ and
$\pi\pi ee$ vertices.  Although these long range contributions have
been analyzed in the form-factor approach~\cite{Vergados:1981bm},
they are more systematically analyzed within the context of EFT
because of the separation of scales:
$m_\pi<\Lambda_{\text{H}}\ll \lbb$.  As noted in
ref.~\cite{Faessler:1996ph}, for example, the matrix elements
associated with  the long range pionic effects
allowed under RPV SUSY scenarios can be dominant.  However, we show
that the dominance of
pion exchange in $\nbb$-decay mediated by heavy physics is a more
general result not limited to RPV SUSY.  These pionic effects can be
considerably larger than those obtained using the conventional
form factor model for the short-range $NNNNee$ process.  For these
reasons, the analysis
of the long range contributions to $\nbb$-decay in EFT will be the main
focus of this paper.

The various types of
$L$-violating operators that contribute to the long range
contributions of $\nbb$-decay appear at different orders in the
$p/ \Lambda_{\text{H}}$ expansion with $p\sim m_\pi$, and the order at
which they appear depends on their symmetry properties.  It is therefore
important to delineate clearly the symmetry properties of
${\cal{L}}_{\text{EFF}}^{\nbb}$ for various types of
$L$-violating operators and use these symmetries to relate the
hadron-lepton operators to the underlying quark-lepton
operators. Carrying out this classification constitutes the first
component of this study. In doing so, we also comment on
the standard approach to deriving $\nbb$-decay nuclear operators and
correct some errors appearing in the literature.

The second step in our treatment involves deriving $\nbb$-decay
nuclear operators from ${\cal{L}}_{\text{EFF}}^{\nbb}$ and expressing
the rate in terms of corresponding nuclear matrix elements.
For any $\beta\beta$-decay mode to occur, the final nucleus must be more
bound than any other
prospective single $\beta$-decay daughter nucleus. Such $\beta$-forbidden
but $\beta\beta$-allowed
nuclei only occur for sufficiently heavy nuclei. Thus, the extraction
of the short-distance physics that gives rise to $\nbb$-decay
(at present, only upper limits on the decay rates exist) depends on a
proper treatment
of the many-body nuclear physics. Having in hand the appropriate set
of nuclear operators (for a given $L$-violation scenario), one could
in principle compute the relevant nuclear matrix elements. Unfortunately,
it is not yet possible to do
so in a manner fully consistent with EFT. This problem has been studied
extensively in the case of the
$NN$ and three-nucleon systems, where the state-of-the art involves use of
chiral symmetry to organize (and
renormalize) the relevant nuclear operators
\cite{vanKolck:yi,Friar:1998zt,bedaque02,beane01}.  Out of necessity, we
follow the same
philosophy here.  Nonetheless, the organization of various
$\nbb$-decay operators based on symmetry considerations and EFT power
counting should represent an
improvement over present treatments of the nuclear problem.

As a final step, we relate the various nuclear operators obtained from
${\cal{L}}_{\text{EFF}}^{\nbb}$ to different particle physics models for
L-violation.  Doing so allows us to determine which nuclear
mechanisms dominate the rate for a given particle physics model.  For
example, in both the RPV SUSY and the left-right symmetric model with
mixing of the gauge bosons, the $\pi\pi ee$ contribution to the
$\nbb$-decay amplitude is
significantly larger  than that of the short range $NNNNee$
contribution.  In contrast, for left-right symmetric models with no
mixing, these contributions are of a similar magnitude.  We also show
how this large $\pi\pi ee$ 
contribution to $\nbb$-decay substantially affects the relationship
between the masses of the right-handed neutrino and gauge boson
including a new correlation between the minimum mass of the
right-handed neutrino and the $W_{\text{L}}-W_{\text{R}}$ mixing
angle.  In short, the 
sensitivity of the $\nbb$-decay searches is strongly affected by
the symmetry transformation properties of the operators contained in a
given particle physics model.

The remainder of our paper is organized as follows. In Section \ref{sec2},
we classify the operators in
${\cal{L}}_{\text{EFF}}^{\nbb}$ according to their symmetry properties and
$p/\Lambda$ counting and we tabulate the various quark-lepton operators
according to the hadron lepton operators they can generate. In Section
\ref{sec3} we use
the leading operators to derive non-relativistic
nuclear operators and compare
their structure with those appearing in conventional treatments. In section
\ref{sec4} we work out the particle physics
implications under various scenarios, namely RPV SUSY and the left-right
symmetric model and compare them to each other.  We
summarize our conclusions in Section \ref{sec5}.

\section{Effective $\nbb$-decay Operators}\label{sec2}

\suppressfloats[t]
\begin{center}
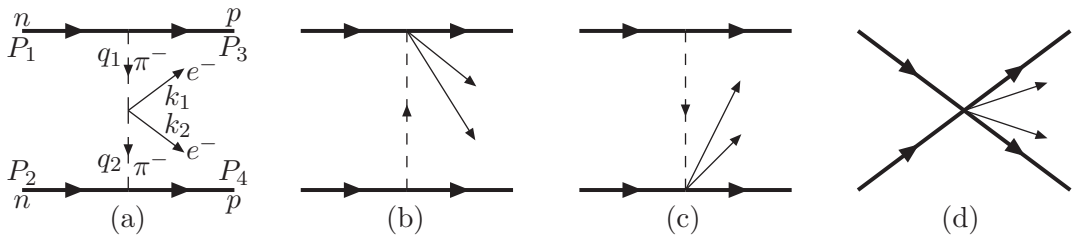
\begin{figure}\caption{Diagrams that contribute to
$\nbb$ at tree level. The exchange diagrams are not
included.}\label{longrangediag}
\begin{picture}(400,80)(0,0)
%
%
{\SetWidth{1.5}
\ArrowLine(5,70)(45,70)
\ArrowLine(45,70)(85,70)
\ArrowLine(5,10)(45,10)
\ArrowLine(45,10)(85,10)}
\Text(5,68)[t]{$P_1$}
\Text(5,12)[b]{$P_2$}
\Text(85,68)[t]{$P_3$}
\Text(85,12)[b]{$P_4$}
\Text(64,50)[t]{$k_1$}
\Text(64,30)[b]{$k_2$}
\Text(43,60)[r]{$q_1$}
\Text(43,20)[r]{$q_2$}
\Text(5,72)[b]{$n$}
\Text(5,8)[t]{$n$}
\Text(85,72)[b]{$p$}
\Text(85,8)[t]{$p$}
\Text(47,60)[l]{$\pi^-$}
\Text(47,20)[l]{$\pi^-$}
\Text(67,56)[l]{$e^-$}
\Text(67,26)[l]{$e^-$}
\DashArrowLine(45,70)(45,40){4}
\DashArrowLine(45,40)(45,10){4}
\Text(45,4)[t]{$\text{(a)}$}
\LongArrow(45,40)(65,55)
\LongArrow(45,40)(65,25)

{\SetWidth{1.5}
\ArrowLine(110,70)(150,70)
\ArrowLine(150,70)(190,70)
\ArrowLine(110,10)(150,10)
\ArrowLine(150,10)(190,10)}
\Text(150,4)[t]{$\text{(b)}$}
\DashArrowLine(150,10)(150,70){4}
\LongArrow(150,70)(175,50)
\LongArrow(150,70)(175,30)

{\SetWidth{1.5}
\ArrowLine(215,70)(255,70)
\ArrowLine(255,70)(295,70)
\ArrowLine(215,10)(255,10)
\ArrowLine(255,10)(295,10)}
\Text(255,4)[t]{$\text{(c)}$}
\DashArrowLine(255,70)(255,10){4}
\LongArrow(255,10)(275,30)
\LongArrow(255,10)(275,50)

{\SetWidth{1.5}
\ArrowLine(320,70)(360,40)
\ArrowLine(360,40)(400,70)
\ArrowLine(320,10)(360,40)
\ArrowLine(360,40)(400,10)}
\Text(360,4)[t]{$\text{(d)}$}
\LongArrow(360,40)(390,50)
\LongArrow(360,40)(390,30)
\end{picture}
\end{figure}
\end{center}
%


The classification of the operators in
${\cal{L}}_{\text{EFF}}^{\nbb}$ relies on two elements:

\begin{itemize}

\item[1.] The use of symmetry to relate effective
lepton-hadron $\nbb$-decay operators to those involving quarks and
leptons.  The relevant symmetries are parity and chiral SU(2).  Indeed,
because the
lepton-hadron effective operators are generated from the quark-lepton
operators through strong interactions, they should retain the
same parity and chiral structure.

\item[2.] The organization of these effective lepton-hadron operators in
  an expansion in powers of a small momentum $p$.

\end{itemize}

To organize the non-standard model (NSM) operators in powers of $p$,
consider first the long range $\pi$-exchange
contributions to $\nbb$-decay of Figs.~\ref{longrangediag}a,b, and c.
The fact that pions are Goldstone bosons
allows us to use chiral perturbation theory
\cite{Gasser:1983yg,Gasser:1984gg} to
classify the NSM hadronic operators in terms of
a $p/\lh$ expansion, with $\lh=4\pi f_\pi\sim 1$ GeV and $p \sim
m_\pi$ where $f_\pi\simeq 92.4$~MeV is the pion decay constant.  The
leading order (LO) quark operators should therefore induce effective
hadronic operators that do not involve derivatives of the pion fields
or pion mass insertions\footnote{At tree level, the pion mass insertions
  always have the form $m_\pi^2$  and therefore  do not contribute at LO or
  NLO.}, the next-to-leading order (NLO) operators would
involve a single derivative of the pion field, the
next-to-next-to-leading order (NNLO) would involve two derivatives or
pion mass insertions and so-on.  This
approach to $\nbb$-decay is similar to the application of effective
field theory to purely hadronic $\Delta S=0$ parity-violating
operators that was done in \cite{Kaplan:1992vj} and the same notation
will be used.

The power counting for the long-range $\nbb$-decay operators will
involve the chiral order of
the standard model (SM) operators as well as the chiral order of the
NSM operators.
For the SM operators, these counting rules are as follows:
\begin{itemize}
\item a pion propagator is  ${\cal{O}}(1/p^2)$ while
\item each derivative of the pion field and the LO strong $\pi NN$
  vertex are  ${\cal{O}}(p)$.
\end{itemize}

As for the short range operators (Fig.~\ref{longrangediag}d, the
hadronic part is constructed
from a 4-nucleon vertex.  This vertex can also be expanded in
powers of the nucleon's 3-momentum.  However, the chiral counting
suggests that the leading ${\cal{O}}(p^0)$ four-nucleon vertex
is already strongly suppressed relative to the long range $\nbb$-decay
operators such that the 4-nucleon vertex can be neglected
to lowest order.  Indeed, with these rules, the chiral counting of the
$\nbb$-decay operators of Figs.~\ref{longrangediag}a-d are
\begin{eqnarray}\label{pcounting}
\text{Fig.~\ref{longrangediag}a}\sim K_{\pi\pi} p^{-2},
~~~ \text{Fig.~\ref{longrangediag}b,c} \sim K_{NN\pi}p^{-1},
~~~ \text{Fig.~\ref{longrangediag}d} \sim K_{NNNN}p^0,
\end{eqnarray}
where the $K_i$ denote the order of the NSM hadronic vertices. In general,
the LO vertex in each
diagram is ${\cal O}(p^0)$, though in certain cases symmetry considerations
require that the leading
order vertex vanish (see below).
Thus, the long range $\nbb$-decay operators of
Figs.~\ref{longrangediag}a, and
\ref{longrangediag}b,c are
enhanced by $1/p^2$ and $1/p$, respectively, relative to the short-range
operator of Fig.~\ref{longrangediag}d.
In what follows, we will consider contributions generated by all of the
diagrams
in Fig.~\ref{longrangediag}. Since the LO contribution from
Fig.~\ref{longrangediag}d
is ${\cal O}(p^0)$, we must include contributions from
Fig.~\ref{longrangediag}a-c
through this order as well. Consequently, we consider all
terms in $K_{\pi\pi}$ and $K_{NN\pi}$  to ${\cal{O}}(p^2)$ and
${\cal{O}}(p)$, respectively.

\subsection{Quark-Lepton Lagrangian}

In order to construct the hadron-lepton operators, we begin by
writing down the quark-lepton Lagrangian for $\nbb$-decay.
This is done by considering all the non-vanishing, inequivalent,
lowest-dimension quark-lepton operators that are
Lorentz-invariant and change lepton number by two units,
\begin{eqnarray}\label{fullquarklag}
{\cal{L}}^{q}_{\nbb} &=&
\frac{G_{\text{F}}^2}{\Lambda_{\beta\beta}}
\left\{
\left(
o_1{\cal{O}}_{1+}^{++}
 +
o_2 {\cal{O}}_{2+}^{++}
 + o_3 {\cal{O}}_{2-}^{++}
+ o_4 {\cal{O}}_{3+}^{++}
+ o_5 {\cal{O}}_{3-}^{++}
\right) \bar{e}e^c
\right.
\nonumber \\
& &
~~~~~~+\left(
o_6{\cal{O}}_{1+}^{++}
 +
o_7 {\cal{O}}_{2+}^{++}
 + o_8 {\cal{O}}_{2-}^{++}
+ o_9{\cal{O}}_{3+}^{++}
+ o_{10} {\cal{O}}_{3-}^{++}
\right) \bar{e}\gamma^5 e^c
\nonumber \\
& &
\left.
~~~~~~+ \left(o_{11} {\cal{O}}_{4+}^{++,\mu} +
o_{12} {\cal{O}}_{4-}^{++,\mu} + o_{13} {\cal{O}}_{5+}^{++,\mu} +
o_{14} {\cal{O}}_{5-}^{++,\mu} \right)\bar{e}\gamma_\mu \gamma^5 e^c
 + \text{h.c.}\right\},
\end{eqnarray}
where
\begin{eqnarray}\label{LOquarkop1}
{\cal{O}}_{1+}^{ab}&=&(\bar{q}_{\text{L}} \tau^a \gamma^\mu
q_{\text{L}})(\bar{q}_{\text{R}} \tau^b \gamma_\mu q_{\text{R}}),
\\
\label{LOquarkop2}
{\cal{O}}_{2\pm}^{ab}&=&(\bar{q}_{\text{R}}\tau^a q_{\text{L}})
(\bar{q}_{\text{R}}\tau^b q_{\text{L}})\pm(\bar{q}_{\text{L}}\tau^a
q_{\text{R}}) (\bar{q}_{\text{L}}\tau^b q_{\text{R}}),
\\
\label{LOquarkop3}
{\cal{O}}_{3\pm}^{ab} &=& (\bar{q}_{\text{L}} \tau^a \gamma^\mu
q_{\text{L}}) (\bar{q}_{\text{L}} \tau^b \gamma_\mu q_{\text{L}})
\pm(\bar{q}_{\text{R}} \tau^a \gamma^\mu
q_{\text{R}}) (\bar{q}_{\text{R}} \tau^b \gamma_\mu q_{\text{R}}),
\\
\label{LOquarkop4}
{\cal{O}}_{4\pm}^{ab,\mu}&=&(\bar{q}_{\text{L}} \tau^a \gamma^\mu
q_{\text{L}} \mp  \bar{q}_{\text{R}}\tau^a \gamma^\mu
q_{\text{R}} ) (\bar{q}_{\text{L}}\tau^b
q_{\text{R}}-\bar{q}_{\text{R}}\tau^b q_{\text{L}}),
\\
\label{LOquarkop5}
{\cal{O}}_{5\pm}^{ab,\mu} &=& (\bar{q}_{\text{L}} \tau^a \gamma^\mu
q_{\text{L}} \pm  \bar{q}_{\text{R}}\tau^a \gamma^\mu
q_{\text{R}} ) (\bar{q}_{\text{L}}\tau^b
q_{\text{R}} + \bar{q}_{\text{R}}\tau^b q_{\text{L}}).
\end{eqnarray}
The $q_{\text{L,R}}=(u,d)_{\text{L,R}}$ are
left-handed and right-handed isospinors and the $\tau$'s are Pauli
matrices in isospace.  When $a=b$, the operators with subscript +(-)
are even (odd) eigenstates of parity as
can be verified by noting that the parity operator simply interchanges
left-handed spinors with right-handed spinors.  This list of nine
operators was arrived at by inspection\footnote{In writing down the
  Eqs.~(\ref{LOquarkop1}-\ref{LOquarkop5}), we suppressed the color
  indices since EFT only relates color-singlet quark
operators to hadronic operators.}.  Other
operators that could have been written down are either equivalent to
those in {Eqs.~(\ref{LOquarkop1})~to~(\ref{LOquarkop5})} or vanish as shown in
appendix~\ref{otherquarkops}.  In particular, all operators
proportional to $\bar{e}\sigma^{\mu\nu}e^c$,
$\bar{e}\gamma^5\sigma^{\mu\nu}e^c$ and $\bar{e}\gamma^\mu e^c$ vanish
since these leptonic currents are identically zero as can
be verified with the use of Fierz transformations.  Some of these vanishing
leptonic currents were erroneously taken as non-zero in
Ref.~\cite{Pas:2000vn}.  Similarly, a quark operator, like
$\bar{q}\sigma^{\mu\nu}\tau^\pm q\bar{q}\sigma_{\mu\nu}\tau^\pm q$, can
be re-expressed in terms of ${\cal{O}}_{2\pm}^{\pm\pm}$ by applying a
Fierz transformation despite the
color indices since the hadronic matrix elements of
four-quark operators only select their color
singlet part\footnote{The projection onto color singlet states
  introduces a new factor that can ultimately be absorbed in the $o_i$'s.}.

Recalling that fermion fields have mass dimension 3/2, note that
the operators appearing in ${\cal{L}}^{q}_{\nbb}$ have
mass dimension nine. Therefore, the overall coefficients have dimensions
$[\text{Mass}]^{-5}$.  In
\Eq{fullquarklag}, this scale factor is expressed as
$G_{\text{F}}^2/\lbb$ where $\lbb$ remains to
be determined.  Derivative quark operators are suppressed by
extra powers of $\lbb$ and need not be considered further.

The operators in ${\cal{L}}^{q}_{\nbb}$ can be generated by various
particle physics models,
but not all of them  are necessarily generated in a single model.
For example, the left-right symmetric model always involves the
product of left-handed and/or right-handed currents, while only
${\cal{O}}_{1+}^{ab}$ and ${\cal{O}}_{3\pm  }^{ab}$ are of that form.
Thus, ${\cal{O}}_{2\pm}^{ab},~{\cal{O}}_{4\pm  }^{ab,\mu}$ and
${\cal{O}}_{5\pm}^{ab,\mu}$ cannot appear in the
left-right symmetric model.  Another example is a
minimal extension of the standard model with only left-handed currents
and Majorana neutrinos; in this scenario, only ${\cal{O}}_{3\pm  }^{ab}$
could appear.  On the other hand, these operators all appear in
RPV SUSY.  This observation will allow a classification of these
particle physics models later in this paper.

Since $\nbb$-decay always requires $a=b=\pm$, the ${\cal{O}}$'s have
definite transformation properties .  Using the quark field
transformation properties under chiral SU(2),
\begin{eqnarray}\label{quarksu2prop}
& &\text{under SU(2$)_{\text{L}}\times$SU(2$)_{\text{R}}$:} ~~~
q_{\text{L}}\rightarrow L q_{\text{L}},~~
q_{\text{R}}\rightarrow R q_{\text{R}},
\end{eqnarray}
where the $L$ and $R$ transformation matrices have the form
exp$\{P_{L,R}\theta_{L,R}\}$ and
\begin{eqnarray}\label{lrmatrices}
\theta_{L,R}\equiv \frac{1}{2}\vec{\tau}\cdot\vec{\theta}_{L,R},~~~~~
P_{L,R} \equiv \frac{1}{2}(1\mp\gamma^5),
\end{eqnarray}
we derive the transformation properties of the
${\cal{O}}_{i\pm}^{ab(\mu)}$  under chiral SU(2),
\begin{eqnarray}\label{SU2quarkop1}
{\cal{O}}_{1+}^{ab}&\rightarrow&(\bar{q}_{\text{L}}L^\dagger \tau^a \gamma^\mu
 L q_{\text{L}})(\bar{q}_{\text{R}} R^\dagger \tau^b \gamma_\mu R
 q_{\text{R}}),
\\
\label{SU2quarkop2}
{\cal{O}}_{2\pm}^{ab}&\rightarrow&(\bar{q}_{\text{R}}R^\dagger \tau^a
L q_{\text{L}})
(\bar{q}_{\text{R}}R^\dagger \tau^b L
q_{\text{L}})\pm(\bar{q}_{\text{L}}L^\dagger\tau^a
R q_{\text{R}}) (\bar{q}_{\text{L}}L^\dagger\tau^b R q_{\text{R}}),
\\
\label{SU2quarkop3}
{\cal{O}}_{3\pm  }^{ab} &\rightarrow& (\bar{q}_{\text{L}} L^\dagger
\tau^a \gamma^\mu
L q_{\text{L}}) (\bar{q}_{\text{L}} L^\dagger \tau^b \gamma_\mu L
q_{\text{L}})
\pm(\bar{q}_{\text{R}} R^\dagger \tau^a \gamma^\mu
R q_{\text{R}}) (\bar{q}_{\text{R}} R^\dagger \tau^b \gamma_\mu R
q_{\text{R}}),
\\
\label{SU2quarkop4}
{\cal{O}}_{4\pm  }^{ab,\mu}&\rightarrow&(\bar{q}_{\text{L}}L^\dagger
\tau^a \gamma^\mu
L q_{\text{L}} \mp  \bar{q}_{\text{R}}R^\dagger \tau^a \gamma^\mu
R q_{\text{R}} ) (\bar{q}_{\text{L}}L^\dagger \tau^b
R q_{\text{R}}-\bar{q}_{\text{R}}R^\dagger \tau^b L q_{\text{L}}),
\\
\label{SU2quarkop5}
{\cal{O}}_{5\pm}^{ab,\mu} &\rightarrow& (\bar{q}_{\text{L}} L^\dagger
\tau^a \gamma^\mu
L q_{\text{L}} \pm  \bar{q}_{\text{R}}R^\dagger \tau^a \gamma^\mu
R q_{\text{R}} ) (\bar{q}_{\text{L}}L^\dagger \tau^b
R q_{\text{R}} + \bar{q}_{\text{R}}R^\dagger \tau^b L q_{\text{L}}).
\end{eqnarray}
We observe that ${\cal{O}}_{1+}^{ab}$
belongs to the $(3_L,3_R)$  representation of
SU(2$)_{\text{L}}\times$SU(2$)_{\text{R}}$ (from here on, the
subscripts $L,R$ are dropped) in the sense
that the first superscript $a$ transforms like a triplet under
SU(2$)_{\text{L}}$ while the second superscript $b$ transforms like a
triplet under SU(2$)_{\text{R}}$.  Note that only
${\cal{O}}_{1+}^{ab}$ belongs to a representation of chiral SU(2).
The other ${\cal{O}}_{i\pm}^{ab(,\mu)}$'s  are
superpositions of operators that have different transformation
properties under chiral SU(2).  This is not surprising since the
generators of chiral SU(2) do not commute with the
parity operator as they involve $\gamma^5$.  For instance,
$(\bar{q}_{\text{L}}
\tau^\pm \gamma^\mu  q_{\text{L}})(\bar{q}_{\text{L}}
\tau^\pm \gamma_\mu  q_{\text{L}})$ changes isospin by two units and
is a singlet under SU(2$)_{\text{R}}$ such that it belongs
to (5,1) while $(\bar{q}_{\text{R}}
\tau^\pm \gamma^\mu  q_{\text{R}})(\bar{q}_{\text{R}}
\tau^\pm \gamma_\mu  q_{\text{R}})$ belongs to (1,5).  Hence,
${\cal{O}}_{3\pm  }^{\pm\pm}$ belongs to  $(5,1)\oplus (1,5)$.

\subsection{Hadron-Lepton Lagrangian}

Let us now turn to the
derivation of the $\pi\pi ee$ vertex from the quark operators.  This
will be followed by a similar analysis for the
$NN\pi ee$ and $NNNNee$ vertices.

\subsubsection{$\pi\pi ee$ vertex.}

To derive the hadronic vertex,
first consider parity.  The product of
two pion fields being even under parity, only positive parity
operators can contribute.  Secondly, note that
${\cal{O}}_{4+  }^{\pm\pm,\mu}$ and ${\cal{O}}_{5+}^{\pm\pm,\mu}$ must
give rise to an operator of the form
\begin{eqnarray}\label{nlopipieeop}
\pi^+\partial^\mu\pi^+\bar{e}\gamma_\mu \gamma^5 e^c +\text{h.c.}
\end{eqnarray}
A partial integration shows that this operator is suppressed by
one power of the electron mass, and is therefore negligible.

Thus, the only terms in ${\cal{L}}^{q}_{\nbb}$ that contribute are:
\begin{eqnarray}\label{pipiquarklag}
\frac{\gf^2}{\lbb}\left\{ {\cal{O}}_{1+}^{++}
\bar{e}(o_1 + o_6 \gamma^5)e^c +
{\cal{O}}_{2+}^{++}\bar{e}(o_2 + o_7 \gamma^5)e^c
+{\cal{O}}_{3+}^{++}
\bar{e}(o_4 + o_{9} \gamma^5)e^c +\text{h.c.}\right\}~.
\end{eqnarray}

The hadronic operators that stem from these
quark operators must
have the same transformation properties and can
be written down by introducing the following fields~\cite{Kaplan:1992vj}:
\begin{eqnarray}\label{fielddef}
\text{X}^a_{\text{R}} &=& \xi\tau^a\xi^\dagger,~~~~
\text{X}^a_{\text{L}} = \xi^\dagger\tau^a\xi, ~~~~ \text{X}^a =
\xi\tau^a\xi, \\ \label{fielddef2}
\xi &=& \text{exp}(i\pi/f_\pi) =
\text{exp}\left[\frac{i}{\sqrt{2}f_\pi}
\left(\tau^+\pi^+ + \tau^-\pi^- + \frac{1}{\sqrt{2}}\tau^3\pi^0\right)
\right]\\ \label{fielddef3}
\pi^{\pm}&=&\frac{1}{\sqrt{2}}(\pi^1\mp i\pi^2), ~~~~~ N:\text{Nucleon
  field.}
\end{eqnarray}
The transformation property of the above fields under parity are
\begin{eqnarray}\label{hadparitytrans}
\pi\rightarrow - \pi, ~~\xi \leftrightarrow \xi^\dagger,~~
\text{X}^a_{\text{R}} \leftrightarrow \text{X}^a_{\text{L}},~~
\text{X}^a \leftrightarrow \text{X}^{\dagger a}, ~~
N\rightarrow \gamma^0 N,
\end{eqnarray}
while under SU(2$)_{\text{L}}\times$SU(2$)_{\text{R}}$ they transform
as
\begin{eqnarray}\label{hadsu2trans1}
\xi &\rightarrow& L\xi U^\dagger = U \xi R^\dagger \\
\label{hadsu2trans2}
X^a  &\rightarrow& U \xi~ R^\dagger\!\tau^a\! L~ \xi U^\dagger \\
\label{hadsu2trans3}
X^a_{\text{L}} &\rightarrow& U \xi^\dagger~ L^\dagger\!\tau^a\! L~ \xi
U^\dagger \\
\label{hadsu2trans4}
X^a_{\text{R}}  &\rightarrow& U \xi~ R^\dagger\!\tau^a\! R~ \xi^\dagger
U^\dagger
\\
\label{hadsu2trans5}
N &\rightarrow& U N~.
\end{eqnarray}
The transformation matrix $U$ only depends on the $\tau$'s
and the pion field.

At LO (no derivatives), the two-pion operator
stemming from the ${\cal{O}}_{1+}^{\pm\pm}$ operator is
\begin{eqnarray}\label{tracephi1}
{\cal{O}}_{1+}^{\pm\pm} \to \text{tr}[\Phi^{\pm\pm}_{1+}] \equiv
   \text{tr}[X_{\text{L}}^\pm X_{\text{R}}^\pm +
X_{\text{R}}^\pm X_{\text{L}}^\pm] =\frac{4}{f_\pi^2}\pi^\mp\pi^\mp
+\cdots,
\end{eqnarray}
while the one generated by ${\cal{O}}_{2+}^{\pm}$
is
\begin{eqnarray}\label{tracephi2}
{\cal{O}}_{2+}^{\pm}\to \text{tr}[\Phi^{\pm\pm}_{2+}] \equiv
   \text{tr}[X^{\pm} X^{\pm} + X^{\dagger\pm}
   X^{\dagger\pm}] = -\frac{4}{f_\pi^2}\pi^\mp\pi^\mp
+\cdots ~~~.
\end{eqnarray}
Here, $\Phi^{\pm\pm}_{1,2\pm }$ are defined
\begin{eqnarray}\label{phi12defs}
\Phi^{\pm\pm}_{1\pm } &\equiv&
  X_{\text{L}}^\pm X_{\text{R}}^\pm \pm
X_{\text{R}}^\pm X_{\text{L}}^\pm~,
\nonumber \\
\Phi^{\pm\pm}_{2\pm } &\equiv&
   X^{\pm} X^{\pm} \pm  X^{\dagger\pm}
   X^{\dagger\pm}~,
\end{eqnarray}
and   the $\pm$ subscript refers to the transformation properties of
the $\Phi_{i\pm}^{\pm\pm}$'s under parity.

Note that when the traces of $\Phi^{\pm\pm}_{1+}$ and
 $\Phi^{\pm\pm}_{2+}$ are
 expanded up to two powers of the pion field, they are physically
indistinguishable since the relative minus sign can be absorbed in a
operator coefficient referred to as a
low energy constant (LEC).

 Now consider the case of the two-pion operator
generated by ${\cal{O}}_{3+}^{\pm\pm}$; to LO the hadronic operator
should be:
\begin{eqnarray}\label{LOleftleft}
\text{tr}\left[X_{\text{L}}^+X_{\text{L}}^+ +
  X_{\text{R}}^+X_{\text{R}}^+ \right] &=&0~.
\end{eqnarray}
Thus, there exists no $(5,1)\oplus (1,5)$ hadronic operator with no
derivatives.

The LO Lagrangian for the $\pi\pi ee$ vertex is therefore
\begin{eqnarray}\label{loppeelag}
{\cal{L}}_{(0)}^{\pi\pi ee}
& = &\frac{G_{\text{F}}^2}{\lbb}\left\{
\text{tr}[\Phi^{++}_{1+}]\bar{e}(a + b\gamma^5)e^c +
\text{tr}[\Phi^{--}_{1+}]\bar{e}^c(a + b\gamma^5)e
\right.
\nonumber \\
& &
\left.
+ \text{tr}[\Phi^{++}_{2+}]\bar{e}(a^\prime + b^\prime\gamma^5)e^c +
\text{tr}[\Phi^{--}_{2+}]\bar{e}^c(a^\prime + b^\prime\gamma^5)e
\right\}~,
\end{eqnarray}
where {$a,b,a^\prime,b^\prime$} are LEC's.  Note
that although there are nominally four LEC's, once the traces of the
$\Phi^{\pm}_i$'s are expanded, there are in practice only two: $a -
a^\prime ~\text{and}~ b - b^\prime$.

In contrast to the $o_i$'s, the {$a,b,a^\prime,b^\prime$} are
dimensionful.  It is useful to express them
in terms of dimensionless parameters (denoted in this work by Greek
letters) with the aid of a scaling rule.  In a
scaling rule, the hadronic operators are divided by the relevant
scales such that their coefficients are dimensionless and  of a
``natural'' size.  We follow the na{\"\i}ve dimensional
analysis (NDA)
scaling rules given in Ref.~\cite{Manohar:1983md} and modified here to
account for the lepton bilinears\footnote{We neglect electromagnetic
  effects.}:
\begin{equation}\label{georgi}
\left({ \frac{\bar{N}N}{\lh f_\pi^2} }\right)^k
\left({\frac{\partial^\mu}{\Lambda_{\text{H}}}}\right)^l
\left({\frac{\pi}{f_\pi}}\right)^m
\left(\frac{f_\pi^2\gf^2}{\lbb} \bar{e}e^c \right)
\ \ \times (\Lambda_{\text{H}} f_\pi)^2.
\end{equation}
Justification for  this scaling rule is given in Appendix~\ref{nda}.
Note that the scaling factor $(\pi/f_\pi)^m$ is already properly
accounted for in the definition of $\xi$ and need not be applied again
in \Eq{loppeelag} after expanding the $\Phi$'s to two pions.  For the
non-derivative $\pi\pi ee$ vertex, we have $(k,l,m)=(0,0,2)$ and
\begin{eqnarray}\label{dimloppeelag}
{\cal{L}}_{(0)}^{\pi\pi ee}
& = &\frac{G_{\text{F}}^2\Lambda_{\text{H}}^2
  f_\pi^2}{\lbb}\left\{
\pi^-\pi^-\bar{e}(\beta_1 + \beta_2\gamma^5)e^c +
\pi^+\pi^+\bar{e}^c(\beta_1 - \beta_2 \gamma^5)e
\right\}.
\end{eqnarray}

Consider now the  higher order contributions to the $\pi\pi ee$
vertex.  As discussed below \Eq{nlopipieeop}, there is no NLO
contribution.  Hence, ${\cal{L}}_{(1)}^{\pi\pi ee} = 0$.

At NNLO, not only do ${\cal{O}}^{\pm\pm}_{1+}$ and
${\cal{O}}^{\pm\pm}_{2+}$ generate two-derivative hadronic operator,
but so does ${\cal{O}}^{\pm\pm}_{3+}$
\begin{eqnarray}\label{nloleftleft}
{\cal{O}}^{\pm\pm}_{3+}\to\frac{1}{2}\text{tr}\left[{\cal{D}}^\mu
X_{\text{L}}^{\pm} {\cal{D}}_\mu
  X_{\text{L}}^{\pm} +
  {\cal{D}}^\mu X_{\text{R}}^{\pm}{\cal{D}}_\mu X_{\text{R}}^{\pm} \right],
\end{eqnarray}
where the chiral covariant derivative is given by
\begin{eqnarray}
{\cal{D}}_\mu=\partial_\mu - i {\cal{V}}_\mu,~~~~~~
{\cal{V}}_\mu=\frac{1}{2}i
\left(\xi\partial_\mu\xi^\dagger + \xi^\dagger\partial_\mu\xi\right).
\end{eqnarray}
The operator ${\cal{D}}_\mu X_{\text{L,R}}$ has the same
transformation properties under chiral SU(2) as $X_{\text{L,R}}$.

The only
other contribution stems from quark mass insertions that always
generate squared pion mass insertions.  Writing the NNLO contributions
directly in terms of pion fields, we obtain
\begin{eqnarray}\label{nnloppeelag}
{\cal{L}}_{(2)}^{\pi\pi ee}
& = &\frac{G_{\text{F}}^2 f_\pi^2}{\lbb}
\left\{
\partial_\mu\pi^- \partial^\mu \pi^-
\bar{e}(\beta_3 + \beta_4\gamma^5)e^c +
m_\pi^2 \pi^-\pi^- \bar{e}(\beta_5 + \beta_6 \gamma^5)e^c
+ \text{h.c.}
\right\}.
\end{eqnarray}
Note that the $\beta_{5,6}$ terms constitute corrections to
$\beta_{1,2}\to \beta_{1,2}+ m_\pi^2\beta_{5,6} $ that can be ignored
in particle physics models where the
LO operators contribute since $\beta_{1,2}$ must be measured\footnote{As
 discussed in Ref.~\cite{Savage:1998yh}, EFT relates the two-derivative
 $\pi\pi ee$ operator to the {\bf 27}-plet $K\to 2\pi$ decays indicating
 the possible existence of an extra suppression factor beyond that
 deduced from power counting.}.

\subsubsection{$NN\pi ee$ vertex}

We analyze the $NN\pi ee$ vertex of
Figs.~\ref{longrangediag}b and \ref{longrangediag}c using similar logic as
in the foregoing discussion.
The LO Lorentz-scalar $NN\pi$
operator is $\bar{N}\tau^\pm \pi^\mp N$ which is odd under parity.
Therefore, ${\cal{O}}_{1+}^{\pm\pm},~{\cal{O}}_{2+}^{\pm\pm}$ and
${\cal{O}}_{3+}^{\pm\pm}$ cannot contribute since they
are parity even.  As for ${\cal{O}}_{3-}^{\pm\pm}$, notice that as in the
$\pi\pi ee$ case, the LO contribution $(X_{\text{L}}^\pm
X_{\text{L}}^\pm - X_{\text{R}}^\pm X_{\text{R}}^\pm)$ vanishes.

The operator $\bar{N}\tau^\pm \pi^\mp N$ can only be induced by
${\cal{O}}_{2-}^{\pm\pm}$.  The result is
\begin{eqnarray}
{\cal{O}}_{2-}^{\pm\pm}&\to &\bar{N}\Phi^{\pm\pm}_{2-} N.
\end{eqnarray}
It is straightforward to verify that
$\bar{N}\Phi^{\pm\pm}_{2-} N$ transforms precisely like
${\cal{O}}_{2-}^{\pm\pm}$ under
SU(2$)_{\text{L}}\times$SU(2$)_{\text{R}}$.

In addition, ${\cal{O}}_{4\pm  }^{\pm\pm,\mu}$
 and ${\cal{O}}_{5\pm}^{\pm\pm,\mu}$ also generate LO contributions to the
  $NN\pi $ operator,
\begin{eqnarray}\label{nnpieequarklag}
{\cal{O}}_{4+  }^{\pm\pm,\mu},{\cal{O}}_{5+}^{\pm\pm,\mu}
& \to & \bar{N}\gamma^\mu \gamma^5 \Phi^{\pm\pm}_{3-}N,
\nonumber \\
{\cal{O}}_{4-  }^{\pm\pm,\mu},{\cal{O}}_{5-}^{\pm\pm,\mu}
& \to & \bar{N}\gamma^\mu \Phi^{\pm\pm}_{3-}N,
\end{eqnarray}
 where
\begin{eqnarray}
\Phi^{\pm\pm}_{3-} = (X^\pm_{\text{L}} + X^{\pm }_{\text{R}})( X^\pm -
  X^{\pm    \dagger }),
\end{eqnarray}
as can be checked explicitly by considering the transformation
properties under chiral SU(2) and parity.  The $NN\pi ee$ LO
Lagrangian can now be written down,
\begin{eqnarray}\label{nnpieeefflag}
{\cal{L}}^{NN\pi ee}_{(0)} &=&
\frac{G_{\text{F}}^2}{\lbb}\left\{
\bar{N}\Phi^{++}_{2-} N\bar{e}(c + d\gamma^5)e^c
+\bar{N}\gamma^\mu (f_1+f_2\gamma^5) \Phi^{++}_{3-}N\bar{e}
\gamma_\mu\gamma^5 e^c
+\text{h.c.}\right\}
\nonumber
\\
&\cong&
\frac{G_{\text{F}}^2\Lambda_{\text{H}} f_\pi}{\lbb}
\left\{\!
\bar{N}\!\tau^+\pi^-\!N\bar{e}(\zeta_1 + \zeta_2\gamma^5)e^c
+\bar{N}\!\gamma^\mu\!(\zeta_3+\zeta_4\gamma^5)\tau^+\pi^-\!N\bar{e}\!
\gamma_\mu\gamma^5\! e^c
+\text{h.c.}\right\},
\end{eqnarray}
where the $\zeta_i$ are dimensionless LEC's introduced using
\Eq{georgi} with $(k,l,m)=(1,0,1)$ and where we have expanded the
$\Phi$'s to one pion.

At NLO, ${\cal{O}}_{1+}^{\pm\pm}$, ${\cal{O}}_{2+}^{\pm\pm}$,
  ${\cal{O}}_{3-}^{\pm\pm}$ and ${\cal{O}}_{3+}^{\pm\pm}$   contribute
  to the $NN\pi$ operator,
\begin{eqnarray}\label{o1+}
{\cal{O}}_{1+}^{\pm\pm} &\to& \bar{N}\gamma^5 \Phi_{1-}^{\pm\pm} N,
\\
\label{o2+}
{\cal{O}}_{2+}^{\pm\pm} &\to& \bar{N}\gamma^5 \Phi_{2-}^{\pm\pm} N,
\\ \label{o3-}
{\cal{O}}_{3-}^{\pm\pm}&\to&
\bar{N}\left\{\gamma^\mu
\left[X_{\text{L}}^\pm (-i{{\cal{D}}}_\mu X_{\text{L}}^\pm ) -
X_{\text{R}}^\pm (i{{\cal{D}}}_\mu X_{\text{R}}^\pm )\right]\right\}N,
\\ \label{o3+}
{\cal{O}}_{3+}^{\pm\pm}&\to&
\bar{N}\left\{\gamma^\mu\gamma^5
\left[X_{\text{L}}^\pm (-i{{\cal{D}}}_\mu X_{\text{L}}^\pm ) -
  X_{\text{R}}^\pm (i{{\cal{D}}}_\mu X_{\text{R}}^\pm
  )\right]\right\}N.
\end{eqnarray}
 The first thing to note is that a term like $\bar{N}\gamma^5\pi N$ is
 subleading because in the non-relativistic reduction, the $\gamma^5$
 couples small and large components of the nucleon spinors.  Secondly,
 we observe that Eqs.~(\ref{o1+}), (\ref{o2+}) and
 (\ref{o3+}) are
 physically indistinguishable on shell when expanded to one pion and
 to the order we are
  considering, as seen from the equations of motion.  Thirdly,
 \Eq{o3-} is negligible
  even at NLO because the equations of motion can be used to show that
  $\bar{N}\notder\pi N$ is proportional to the electron
  momentum.  Therefore, ${\cal{O}}_{3-}^{\pm\pm}$ does not contribute
  to the $NN\pi ee$ vertex.

Other contributions to
  ${\cal{O}}(p)$ include
terms normally neglected at LO in the non-relativistic reduction of
  \Eq{nnpieeefflag}, namely the terms
proportional to $\zeta_3$ and $\zeta_4$ with $\mu=1,2,3$ and $\mu=0$
respectively, where LO and NLO components of the nucleon spinors are
coupled.  These are the only contributions to the $NN\pi ee$ vertex
since the $m_\pi^2$ insertions are of ${\cal{O}}(p^2)$ and excluded as
discussed below \Eq{pcounting}.  Hence, the only new contributions to
${\cal{O}}(p)$ is,
\begin{eqnarray}\label{expnlonnpiee}
{\cal{L}}^{NN\pi ee}_{(1)} &=&  \frac{\gf^2 \lh f_\pi}{\lbb}
 \bar{N}\!\gamma^5 \tau^+ \pi^-\! N
\bar{e} (\zeta_5 + \zeta_6 \gamma^5) e^c
+ \text{h.c.}
  \end{eqnarray}
where the scaling rule in \Eq{georgi} was used with
$(k,l,m)=(1,1,1)$.  ${\cal{L}}^{NN\pi ee}_{(1)}$ is subleading because
the $\gamma^5$ couples the large and small components of the nucleon
spinors and the result is proportional to $p/M$ where $M$ is the
nucleon mass and $p$ is the magnitude of the nucleon three-momentum.

\subsubsection{$NNNNee$ vertex}

To identify the quark operators that generate the $\nbb$-decay
four-nucleon operators, we  insert the hadronic fields
$X^{\pm\pm}_{\text{LR}},~X^{\pm\pm},~X^{\dagger\pm\pm}$ in all
possible ways into $\bar{N}\Gamma N\bar{N}\Gamma^\prime N$
and use their transformation properties under chiral SU(2) to relate
them to the ${\cal{O}}_{i\pm}^{\pm\pm(,\mu)}$.  The four-nucleon
operators are then obtained by expanding these hadronic fields
to LO and ignoring all contributions from pion loops.  Thus, it is not
necessary to insert these hadronic fields in all possible ways; we
only need to show that a particular quark operator can generate a
particular nucleon operator with the same transformation properties
under parity and chiral SU(2).

For example, the LO operator $(\bar{N}\tau^\pm N)^2$
can be generated by ${\cal{O}}_{1+}^{\pm\pm}$. The latter transforms the
same way under parity and
chiral SU(2) as the hadronic operator
\begin{equation}
\label{eq:firstcontact}
(\bar{N} X_{\text{L}}^\pm N ) (
  \bar{N} X_{\text{R}}^\pm N)\ \ \ .
\end{equation}
At zero pion order, the $X_{L}^{\pm}$ and $X_{R}^{\pm}$ both become
$\tau^{\pm}$, so
that  the operator in Eq. (\ref{eq:firstcontact}) just becomes
$(\bar{N}\tau^\pm N)^2$.
In a similar fashion, it can be easily shown that the following five
operators
\begin{eqnarray}\label{basicnnnn}
{\mathfrak{N}}^{\pm\pm}_{1+} = (\bar{N}\tau^\pm N)^2, ~  ~
{\mathfrak{N}}^{\pm\pm}_{2+} =(\bar{N}\tau^\pm \gamma^\mu
N)(\bar{N}\tau^\pm \gamma_\mu N), ~  ~
{\mathfrak{N}}^{\pm\pm}_{3+} =(\bar{N}\tau^\pm \gamma^5\gamma^\mu
N)(\bar{N}\tau^{\pm}\gamma^5 \gamma_\mu N), ~  ~
& &
\nonumber
\\
{\mathfrak{N}}^{\pm\pm,\mu}_{4+} =(\bar{N}\tau^\pm \gamma^\mu
N)(\bar{N}\tau^\pm N), ~ ~
{\mathfrak{N}}^{\pm\pm,\mu}_{4-} =(\bar{N}\tau^\pm \gamma^5\gamma^\mu
N)(\bar{N}\tau^\pm N),~~~~~& &
\end{eqnarray}
exhaust the list of possible LO four-nucleon
operators\footnote{Since $\bar{N}\gamma^5 N $ and
  $(\bar{N}\gamma^5\gamma^\mu N)( \bar{N}\gamma_\mu N) $ are
  proportional to $p/M$, they are
sub-leading in the non-relativistic limit.} that can be
generated by the checked
${\cal{O}}_{i\pm}^{\pm\pm(,\mu)}$'s in Table~\ref{nnnntable}.
\begin{table}\caption{Cross-reference table between nucleon and quark
    operators.  The $X$ indicates that the quark operator cannot generate
    the corresponding nucleon operator while the $\surd$ indicates
    that it can. } \label{nnnntable}

\begin{tabular}{|c|c|c|c|c|c|c|c|c|c|}\hline
$NNNN$ ops.
 & ${\cal{O}}_{1+}^{\pm\pm}$
 & ${\cal{O}}_{2+}^{\pm\pm}$
 & ${\cal{O}}_{2-}^{\pm\pm}$
 & ${\cal{O}}_{3+}^{\pm\pm}$
 & ${\cal{O}}_{3-}^{\pm\pm}$
 & ${\cal{O}}_{4+ }^{\pm\pm,\mu}$
 & ${\cal{O}}_{4- }^{\pm\pm,\mu}$
 & ${\cal{O}}_{5+}^{\pm\pm,\mu}$
 & ${\cal{O}}_{5-}^{\pm\pm,\mu}$ \\
\hline
${\mathfrak{N}}^{\pm\pm}_{1+}$
& $\surd$
& $\surd$
& $X$
& $\surd$
& $X$
& $X$
& $X$
& $X$
& $X$ \\
\hline
${\mathfrak{N}}^{\pm\pm}_{2+}$
& $\surd$
& $\surd$
& $X$
& $\surd$
& $X$
& $X$
& $X$
& $X$
& $X$ \\
\hline
${\mathfrak{N}}^{\pm\pm}_{3+}$
& $\surd$
& $\surd$
& $X$
& $\surd$
& $X$
& $X$
& $X$
& $X$
& $X$\\
\hline
${\mathfrak{N}}^{\pm\pm,\mu}_{4+}$
& $X$
& $X$
& $X$
& $X$
& $X$
& $\surd$
& $X$
& $\surd$
& $X$\\
\hline
${\mathfrak{N}}^{\pm\pm,\mu}_{4-}$
& $X$
& $X$
& $X$
& $X$
& $X$
& $X$
& $\surd$
& $X$
& $\surd$\\
\hline
\end{tabular}
\end{table}

The LO four-nucleon Lagrangian is therefore given by
\begin{eqnarray}\label{nnnnlag}
{\cal{L}}^{NNNN ee}_0 &=&
\frac{\gf^2}{\lbb}\left\{
\left(\xi_1 \mathfrak{N}_{1+}^{++}  + \xi_2 \mathfrak{N}_{2+}^{++} + \xi_3
\mathfrak{N}_{3+}^{++} \right) \bar{e}e^c
+\left( \xi_4 \mathfrak{N}_{1+}^{++}  + \xi_5 \mathfrak{N}_{2+}^{++} + \xi_6
\mathfrak{N}_{3+}^{++} \right) \bar{e} \gamma^5 e^c
\right.
\nonumber
\\
& &~~~~~~~~~~~~~~~~~~~~~~~~~~~~~~~~~~
\left.
+\left( \xi_7 \mathfrak{N}_{4+}^{++,\mu}  + \xi_8 \mathfrak{N}_{4-}^{++,\mu}
 \right) \bar{e} \gamma^5 \gamma_\mu e^c
+\text{h.c.}
\right\} ,
\end{eqnarray}
where the $\xi_i$'s are dimensionless.

 In concluding this section, we discuss a few issues that will require future
 work. The first involves the application of EFT to heavy nuclei. 
 As pointed out earlier, no fully consistent treatment for such
 situations has yet been developed. In principle,
 one could imagine following a program similar in spirit to the EFT
 treatment of 
 few-body systems. In that case, there has been recent progress in developing
 a consistent power counting for EFT with explicit
 pions\cite{beane01,bedaque02}. The approach involves including the LO
 $\pi$-exchange contribution to the NN potential,
 expanding it about the chiral limit ($m_\pi^2\to 0$), and obtaining
 two-body wavefunctions
 by solving the Schrodinger equation with the chirally-expanded potential.
 To be consistent, operators would also be expanded to the same chiral order
 as the potential and matrix elements computed using the corresponding
 wavefunctions. This approach appears to reproduce the consistent momentum
 power counting obtained with perturbative pions in the $^1S_0$ channel and
 the convergence obtained with non-perturbative pions in the $^3S_1$-$^3D_1$
 channel.
 In going to more complex nuclei, one might explore a marriage of the chiral
 expansion with traditional many-body techniques ({\em e.g.}, shell model or
 RPA), in which case one would require a corresponding chiral counting of
 nuclear operators. In organizing the $0\nu\beta\beta$-decay hadronic
 operators according to both the derivative and chiral expansion, we have
 taken one step in this direction. For the moment, however, we will have to
 content ourselves with using these operators along with  wavefunctions
 obtained from traditional many-body techniques.
 
 A second issue is the presence of higher partial waves in the two-body
 transition matrix elements appearing in $0\nu\beta\beta$-decay.
 A fully consistent treatment would, therefore, require that one
include the corresponding higher-order operators -- a task that is
 clearly impractical at 
present.  Fortunately, in our case, there is reason to believe our qualitative
 conclusions about the dominance
 of long-range, pion-exchange operators are fairly insensitive to this
 issue. For the cases where
 the LO $\pi\pi ee$ are not forbidden by the symmetries of the quark-lepton
operators, the LO $\pi$-exchange operator arising from
 Fig.~\ref{longrangediag}a will always give the LO contribution
 to the transition matrix element, regardless of the partial wave
 decomposition of the two-nucleon
 initial and final states. In general, then,  we expect that matrix
 elements of these 
 operators should always be enhanced relative to those involving the
 four-nucleon contact operators or $\pi$-exchange operators obtained with
 higher-order pionic vertices.  Indeed, some evidence to this effect is
 given by the computation of Ref. \cite{Faessler:1998qv}, where the relative
 importance of the LO $\pi$-exchange operators and short-range operators
 were compared for RPV SUSY\footnote{However, in that work, the
 traditional, form factor approach was used to compute short-range effects.}.

%

Finally, when NNLO and NLO interactions are included at tree level,
loop graphs must also be included to be consistent with the power
counting (examples of which are given in Fig.~\ref{renormgraphs}).
These loop graphs are handled according
to the chiral perturbation theory prescription by which the
divergences renormalize the LEC's that multiply the $m_\pi^2$ and
two-derivative $\pi\pi ee$ vertex of Eq.~(\ref{nnloppeelag}).  In this
context, loop graphs that 
renormalize the $NN\pi ee$ vertex are N$^3$LO and can be ignored.
Indeed, this can be demonstrated using power counting where each loop
involves a factor of $p^4$ while nucleon propagators count as
$p^{-1}$~\cite{Becher:1999he,Jenkins:1990jv,Lehmann:2001xm}.

When loops are included, new lepton-violating tree
level vertices can contribute inside the loop graphs, such as the
$\pi\pi\pi\pi ee$ vertex of \Fig{renormgraphs}b.  Other new vertices
that could potentially contribute at the one loop level are $NN\pi\pi
ee$ and $\pi\pi\pi ee$ vertices.  In short,
large number of Feynman diagrams may need to be calculated at NNLO.  We defer
a discussion of such loop contributions to a subsequent study.
\suppressfloats[t]


\begin{center}
\begin{figure}\caption{ \sl \footnotesize a) Example of a graph that
    renormalizes the LEC's that multiplies the $m_\pi^2$ and
    two-derivative $\pi\pi ee$ 
    vertex.  b)  Example of a new vertex ($\pi\pi\pi\pi ee$) that
    contributes to $\nbb$ at NNLO.}\label{renormgraphs}
\begin{picture}(360,120)(0,0)
%
%
{\SetWidth{1.5}
\ArrowLine(0,90)(80,90)
\ArrowLine(80,90)(160,90)
\ArrowLine(0,4)(80,4)
\ArrowLine(80,4)(160,4)}
\DashArrowLine(80,90)(80,43){4}
\DashArrowLine(80,4)(80,43){4}
\DashCArc(100,47)(20,0,360){4}
\LongArrow(120,47)(160,69)
\LongArrow(120,47)(160,25)

\Text(5,92)[b]{$n$}
\Text(155,92)[b]{$p$}
\Text(5,2)[t]{$n$}
\Text(155,2)[t]{$p$}
\Text(78,67)[r]{$\pi$}
\Text(160,68)[t]{$e^-$}
\Text(160,30)[b]{$e^-$}
\Text(80,0)[t]{(a)}

%
{\SetWidth{1.5}
\ArrowLine(200,90)(280,90)
\ArrowLine(280,90)(360,90)
\ArrowLine(200,4)(280,4)
\ArrowLine(280,4)(360,4)}
\DashArrowLine(280,90)(280,47){4}
\DashArrowLine(280,4)(280,47){4}
\DashCArc(260,47)(20,0,360){4}
\LongArrow(280,47)(320,69)
\LongArrow(280,47)(320,25)

\Text(280,0)[t]{(b)}
\end{picture}
\end{figure}
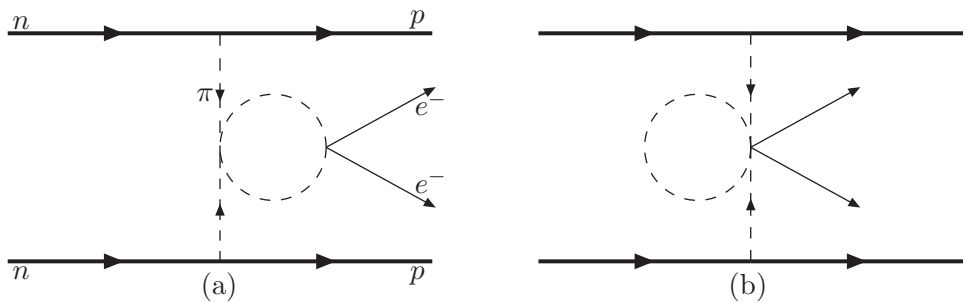
\end{center}

To summarize the conclusions of the analysis, Table~\ref{qqeehheetable}
lists the  quark-lepton operators that contribute to the various
hadron operators at LO.  One important result indicated in the table
is the fact that if the short-distance physics
responsible for $\nbb$-decay belongs to a representation of
SU$(2)_L\times$SU$(2)_R$,
only operators that belong to the (3,3) and
$(5,1)\oplus (1,5)$ can generate $\nbb$-decay and therefore, only
${\cal{O}}_{1+}^{\pm\pm}$ and ${\cal{O}}_{3+  }^{\pm\pm}$ can
contribute.  For example, the left-right symmetric model with
mixing between left- and right-handed gauge bosons induces operators
belonging to the (3,3) as well as the $(5,1)\oplus (1,5)$. From
Table~\ref{qqeehheetable}, the LO $\nbb$-decay
operator that  contributes in this case is generated by
Fig.~\ref{longrangediag}a and is ${\cal O}(p^{-2})$.

Alternatively, consider a
short-distance model involving products of two left-handed
currents or two right-handed currents only. Such a situation arises, for
instance, in the left-right symmetric model
when the $W_L$ and $W_R$ bosons do not mix. For this scenario, only
${\cal{O}}_{3+  }^{\pm\pm}$ contributes, and there are  no LO
contributions to the $\pi\pi ee$ and $NN\pi ee$ vertices.  The first
non-zero contributions to the hadronic part of these vertices are
given by Eqs.~(\ref{nloleftleft}) and (\ref{o3+}) as well as
contributions that include $m_\pi^2$ insertions.  The resulting
contribution to the
amplitude is ${\cal O}(p^0)$. In this case, both the long- and short-range
nuclear operators
occur at the same order.

\begin{table}\caption{Leading order $\nbb$-decay hadronic-lepton
    operators generated
    by the various quark-lepton operators. } \label{qqeehheetable}

\begin{tabular}{|c|c|c|c|c|c|c|c|c|c|}\hline
$\nbb$-decay ops.
 & ${\cal{O}}_{1+}^{\pm\pm}$
 & ${\cal{O}}_{2+}^{\pm\pm}$
 & ${\cal{O}}_{2-}^{\pm\pm}$
 & ${\cal{O}}_{3+}^{\pm\pm}$
 & ${\cal{O}}_{3-}^{\pm\pm}$
 & ${\cal{O}}_{4+ }^{\pm\pm,\mu}$
 & ${\cal{O}}_{4- }^{\pm\pm,\mu}$
 & ${\cal{O}}_{5+}^{\pm\pm,\mu}$
 & ${\cal{O}}_{5-}^{\pm\pm,\mu}$ \\
\hline
$\pi\pi ee$ LO
& $\surd$
& $\surd$
& $X$
& $X$
& $X$
& $X$
& $X$
& $X$
& $X$ \\
\hline
$\pi\pi ee$ NNLO
& $\surd$
& $\surd$
& $X$
& $\surd$
& $X$
& $X$
& $X$
& $X$
& $X$ \\
\hline
$NN\pi ee$ LO
& $X$
& $X$
& $\surd$
& $X$
& $X$
& $\surd$
& $\surd$
& $\surd$
& $\surd$ \\
\hline
$NN\pi ee$ NLO
& $X$
& $\surd$
& $X$
& $\surd$
& $X$
& $\surd$
& $\surd$
& $\surd$
& $\surd$ \\
\hline
$NNNN ee$ LO
& $\surd$
& $\surd$
& $X$
& $\surd$
& $X$
& $\surd$
& $\surd$
& $\surd$
& $\surd$\\
\hline
\end{tabular}
\end{table}

\section{Nuclear Operators to LO and NLO}\label{sec3}

In the calculation of the $\nbb$-decay amplitude, the Feynman diagrams
of \Fig{longrangediag} must be calculated to
${\cal{O}}(p^0)$, where $p$ is the small momentum used as an
expansion parameter.  As discussed below \Eq{pcounting}, this implies
that we need to include NNLO  $\pi\pi ee$
operators, NLO $NN\pi ee$ operators and LO $NNNNee$ operators.

From the $\pi\pi ee$ Lagrangian of \Eq{dimloppeelag}, the LO $\nbb$-decay
amplitude of \Fig{longrangediag}a is calculated to be
\begin{equation}\label{diaga}
M_0^{\pi\pi} = -\frac{g_{\text{A}}^2\gf^2\lh^2 M^2}{\lbb}
\frac{8(\bar{u}_{p3} \gamma^5 u_{n1})
(\bar{u}_{p4}\gamma^5
u_{n2})}{(q_1^2 - m_\pi^2 + i\epsilon)(q_2^2 - m_\pi^2 +
i\epsilon)}
\times \bar{u}_{e1} \gamma^2\gamma^0 (\beta_1 + \beta_2\gamma^5)
\bar{u}_{e2}^{\text{T}}~,
\end{equation}
where $q_1=P_1-P_3,~q_2=P_2-P_4$ as defined in \Fig{longrangediag}a
and $g_{\text{A}}=1.27$ is the usual axial pion-nucleon coupling
related to $g_{\pi NN}$ by the Goldberger-Treiman relation.

As for the NLO, recall from \Eq{nlopipieeop} and the discussion that
followed that the $\pi\pi ee$ vertex has no NLO
contributions.  Thus, the NLO $\nbb$-decay nuclear operators are given by
\Fig{longrangediag}b and \ref{longrangediag}c.  Note that
experiments planned and  under way
involve mainly ground state to ground state transitions $0^+\to 0^+$
which are favored by phase space considerations.  The nuclear matrix elements
of all the operators of ${\cal{L}}_0^{NN\pi ee}$ [\Eq{nnpieeefflag}]
vanish for this transition by parity\footnote{Recall from above that
$\bar{N}\gamma^5 N$ and $\bar{N}\gamma^i N$, $i=1,2,3$, are NNLO
operators that couple the large and small components of the nucleon
spinors.}.  There are therefore no NLO contributions for the $0^+\to 0^+$
transition and $M_0^{\pi\pi}$ is the only non-vanishing amplitude through
${\cal O}(p)$.  Nevertheless, we provide the expressions for the NLO nuclear
operators in Appendix~\ref{appnloops} for completeness.

Taking the non-relativistic limit of \Eq{diaga} and Fourier transforming to
co-ordinate space yields
\begin{eqnarray}\label{approxa}
\text{F.T.}M_0^{\pi\pi} &\simeq&
\frac{1}{12\pi}
\frac{g_{\text{A}}^2\gf^2\lh^2}{\Lambda_{\beta\beta}}
~\bar{u}_{e1} \gamma^2\gamma^0(\beta_1+ \beta_2 \gamma^5)
\bar{u}_{e2}^{\text{T}}
{\cal{O}}_0^{\pi\pi}(\vec{x}_1,\vec{x}_2,\vec{x}_3,\vec{x}_4),
\end{eqnarray}
where the nuclear operator is given by
\begin{eqnarray}\label{lonuclop}
{\cal{O}}_0^{\pi\pi}(\vec{x}_1,\ldots,\vec{x}_4)
=
-\delta(\vec{x}_1-\vec{x}_3) \delta(\vec{x}_2-\vec{x}_4)
(\chi_{3,\alpha}^\dagger  \chi_{1,\beta})
(\chi_{4,\phi}^\dagger\chi_{2,\delta})
\frac{1}{\rho}
[\text{F}_1\vec{\sigma}_{\alpha\beta}\cdot\vec{\sigma}_{\phi\delta}
+\text{F}_2 T_{\alpha\phi,\beta\delta}],
\end{eqnarray}
and
\begin{eqnarray}
T_{\alpha\phi,\beta\delta}\equiv
3\vec{\sigma}_{\alpha\beta}\cdot\hat{\rho}
\vec{\sigma}_{\phi\delta}\cdot\hat{\rho}
- \vec{\sigma}_{\alpha\beta}\cdot\vec{\sigma}_{\phi\delta}~.
\end{eqnarray}
The form-factors $\text{F}_1$ and $\text{F}_2$ were first introduced
in Ref.~\cite{Vergados:1981bm}
\begin{eqnarray}
\text{F}_1(x)=(x-2)\text{e}^{-x},~~~~~~~
\text{F}_2(x)=(x+1)\text{e}^{-x},
\end{eqnarray}
where $x = m_\pi \rho$, $\rho=|\vec{x}_1-\vec{x}_2|$ is the distance
between the nucleons,
and ${\hat\rho}={\vec\rho}/\rho$.
However, in Ref.~\cite{Vergados:1981bm}, these form-factors were
derived within
a minimal extension of the standard model with only left-handed
currents and heavy Majorana neutrinos; as was shown above by
considering the possible representations to which the product of two
left-handed weak currents can belong, this minimal extension cannot
give rise to the LO $\pi\pi ee$ vertex that yields these form-factors.
In contrast, the derivation of $F_1$ and $F_2$ was performed here by
considering the symmetry properties of the quark
operators that could generate the hadronic $\nbb$-decay operators without
specifying the short-distance physics responsible for $\nbb$-decay.

Up to NLO, the $\nbb$-decay half-life is therefore
\begin{eqnarray}\label{zeroderivativerate}
\frac{1}{T_{1/2}} &=&
\frac{\hbar c^2}{144\pi^5\ln\!2}\frac{g_{\text{A}}^4}{R^2}
  \frac{\Lambda_{\text{H}}^4\gf^4}{\Lambda_{\beta\beta}^{2}}
  \int_{m_e}^{E_{\beta\beta} - m_e}\text{d}\!E_1
  F\!(Z+2,E_1)F\!(Z+2,E_2) ~~~~
\nonumber \\
  & & ~~~~~~~~~~~~~~~~~~\frac{1}{2} [(\beta_1^2+\beta^2_2)p_1E_1p_2E_2
  -   (\beta_1^2-\beta^2_2)p_1p_2m_e^2] |{\cal{M}}_0|^2,
\end{eqnarray}
where $F(Z,E)$ is the usual Fermi function describing the Coulomb
effect on the outgoing electrons with
\begin{eqnarray}\label{matrixelement}
{\cal{M}}_0 &=& <\Psi_{A,Z+2}|
\sum\limits_{ij}\frac{R}{\rho_{ij}}
[ F_1\!(x_{ij}) \vec{\sigma}_i \cdot \vec{\sigma}_j
+ F_2\!(x_{ij}) T_{ij} ] \tau^+_i\tau^+_j
 | \Psi_{A,Z} >,
\\
T_{ij} &=& 3\vec{\sigma}_i \cdot \hat{\rho}_{ij}
\vec{\sigma}_j \cdot \hat{\rho}_{ij} - \vec{\sigma}_i \cdot
\vec{\sigma}_j,
\\
E_2 &=& E_{\beta\beta} - E_1,~~~ p_i = \sqrt{E_i^2 - m_e^2}~.
\end{eqnarray}
Here $\rho_{ij}$ is the distance between the $i$'th and $j$'th
neutrons in the initial nucleus $| \Psi_{A,Z} >$ or the distance
between two protons in the final state  $| \Psi_{A,Z+2} >$, $m_e$ is
the mass of
the electron, $R$ is a scale taken to be of the order of the nuclear
radius\footnote{This scale is inserted to make the operator in
\Eq{matrixelement} dimensionless. It is canceled by a corresponding factor
of $1/R^2$ in the rate.} $\vec{\sigma}_{i(j)}$ acts
on the spin of the $i(j)$'th neutron and the isospin matrix
$\tau^+_{i(j)}$ turns the $i(j)$'th neutron into a proton.  Note that
independently of the nuclear matrix element,
the $\beta_1^2-\beta_2^2$ part of the rate in \Eq{zeroderivativerate}
is always considerably smaller (by at least a factor of $\sim 10$ from
the kinematics) than
the  $\beta_1^2+\beta_2^2$ part which is the only one usually considered.

\subsection{Long Range Operators At NNLO}

Consider now the long range operators at NNLO.  We are
interested in comparing the LO and NNLO tree-level long range
contributions and for simplicity we will ignore contributions from
loops, $m_\pi^2$ insertions and the four-nucleon vertex which
also contribute at NNLO\footnote{We also ignore recoil order
  corrections from the amplitude of Fig.~\ref{longrangediag}a where
  $K_{\pi\pi}$ is of ${\cal{O}}(p^0)$.  In this case, the rate will be
dominated by terms in \Eq{zeroderivativerate}.}.
Thus, we only need the hadronic operators of
Eqs.~(\ref{nnloppeelag}) and (\ref{expnlonnpiee}) rewritten here
\begin{eqnarray}\label{smoperators}
M_2=\frac{G_{\text{F}}^2 }{\lbb}
\left\{
f_\pi^2 \partial_\mu\pi^- \partial^\mu \pi^-
\bar{e}(\beta_3 + \beta_4\gamma^5)e^c
+ f_\pi\lh \bar{N}\!\gamma^5 \tau^+ \pi^-\! N
\bar{e} (\zeta_5 + \zeta_6 \gamma^5) e^c
+ \text{h.c.}
\right\}.
\end{eqnarray}
The diagrams of
Fig.~\ref{longrangediag}a,b and c can be evaluated using the operators of
Eq.~(\ref{smoperators}).  The Fourier transform of the final result
is:
\begin{eqnarray}\label{smft}
M_2=
\frac{1}{8\pi}g_{\text{A}}^2 \frac{G_{\text{F}}^2}{\lbb}
\left[
\bar{u}_{e1} \gamma^2\gamma^0(\beta_3 + \beta_4\gamma^5)
\bar{u}_{e2}^{\text{T}}~
{\cal{O}}_2^{\pi\pi}(\vec{x}_1,\ldots ,\vec{x}_4)
 \right.~~~~~~~~~~~~~~~~~
& &
\nonumber \\
\left.
 +
\bar{u}_{e1} \gamma^2\gamma^0(\zeta_5 + \zeta_6\gamma^5)
\bar{u}_{e2}^{\text{T}}~
{\cal{O}}_2^{\pi NN}(\vec{x}_1,\ldots , \vec{x}_4)
 \right],
& &
\end{eqnarray}
with
\begin{eqnarray}\label{nnlonuclop}
{\cal{O}}_2^{\pi\pi}(\vec{x}_1,\ldots , \vec{x}_4)&=&
-\delta(\vec{x}_1-\vec{x}_3) \delta(\vec{x}_2-\vec{x}_4)
(\chi_{3,\alpha}^\dagger  \chi_{1,\beta})
(\chi_{4,\phi}^\dagger\chi_{2,\delta})
\frac{1}{\rho^3}\\
&&\times
\left( \text{G}_1^{\pi\pi}\vec{\sigma}_{\alpha\beta}
  \cdot\vec{\sigma}_{\phi\delta} +
\text{G}_2^{\pi\pi} T_{\alpha\phi,\beta\delta}
\right)\nonumber
\\
{\cal{O}}_2^{\pi NN}(\vec{x}_1,\ldots , \vec{x}_4) &=&
- \frac{\sqrt{2}\lh}{g_{\text{A}}M} \delta(\vec{x}_1-\vec{x}_3)
\delta(\vec{x}_2-\vec{x}_4)
(\chi_{3,\alpha}^\dagger  \chi_{1,\beta})
(\chi_{4,\phi}^\dagger\chi_{2,\delta})
\frac{1}{\rho^3}\\
&&
 \times\left(
\text{G}_1^{\pi NN} \vec{\sigma}_{\alpha\beta}
  \cdot\vec{\sigma}_{\phi\delta} +
   \text{G}_2^{\pi NN} T_{\alpha\phi,\beta\delta}
\right),
\nonumber
\end{eqnarray}
and ($x=m_\pi \rho$ as before)
\begin{eqnarray}
G_1^{\pi\pi} &=& -\frac{x^2}{3}( 4 - x) e^{-x},\\
G_2^{\pi\pi} &=& -\left[2 +2x + \frac{1}{3} x^2 - \frac{1}{3}
x^3 \right] e^{-x},
\\
G_1^{\pi NN} &=& -\frac{1}{3}x^2 e^{-x},\\
G_2^{\pi NN} &=& -(1 + x + \frac{1}{3}x^2 ) e^{-x}.
\end{eqnarray}
The new form-factors $\text{G}_1^{\pi\pi}$ and $\text{G}_2^{\pi\pi}$
stem from the $\pi\pi
ee$ vertex while $\text{G}_1^{\pi NN}$ and $\text{G}_2^{\pi NN}$ (also
given in Ref.~\cite{Faessler:1998qv}) stem
from the $NN\pi ee$ vertex.  In
contrast to the zero-derivative case, the amplitudes stemming from
these two vertices are of the same order in this minimal extension of
the standard model.

The corresponding half-life, assuming that \Eq{smft} represents the
only decay amplitude,
is:
\begin{eqnarray}\label{twoderivativerate}
\frac{1}{T_{1/2}} =
\frac{1}{64\pi^5\ln\!2}\left(\frac{\hbar c}{R}\right)^6
\frac{g_{\text{A}}^4}{\hbar}
  \frac{G_{\text{F}}^4}{\lbb^2 c^4}
~~~~~~~~~~~~~~~~~~~~~~~~~~~~~~~~~~~~~~~~~~~~~~~~~~~~~~~~~~~~~& &
\nonumber \\
\times  \int_{m_e}^{E_{\beta\beta} - m_e}\text{d}\!E_1
  F\!(Z+2,E_1)F\!(Z+2,E_2) ~~~~~~~~~~~~~~~~~~~~~~~~~~~~~~~~~~
\nonumber \\
\frac{1}{2}
\left\{
\left[
\left|\beta_3{\cal{M}}_2^{\pi\pi
  }+\frac{\sqrt{2}\lh}{g_{\text{A}}M}\zeta_5{\cal{M}}_2^{\pi NN
  }\right|^2
+
\left|\beta_4{\cal{M}}_2^{\pi\pi
  }+\frac{\sqrt{2}\lh}{g_{\text{A}}M}\zeta_6{\cal{M}}_2^{\pi NN
  }\right|^2
\right]
p_1E_1p_2E_2
\right.
~~& &
\nonumber \\
\left.
-\left[
\left|\beta_3{\cal{M}}_2^{\pi\pi
  }-\frac{\sqrt{2}\lh}{g_{\text{A}}M}\zeta_5{\cal{M}}_2^{\pi NN
  }\right|^2
-
\left|\beta_4{\cal{M}}_2^{\pi\pi
  }-\frac{\sqrt{2}\lh}{g_{\text{A}}M}\zeta_6{\cal{M}}_2^{\pi NN
  }\right|^2
\right]
p_1p_2m_e^2
\right\}~,
& &
\end{eqnarray}
with
\begin{eqnarray}
{\cal{M}}_2^{\pi\pi (\pi NN)} &=& <\Psi_{A,Z+2}|
\sum\limits_{ij} \left( \frac{R}{\rho_{ij}} \right)^3
 \left[ G_1^{\pi\pi(\pi NN)}\!(x_{ij})
\vec{\sigma}_i \cdot \vec{\sigma}_j
\right.
\nonumber \\
& & ~~~~~~~~~~~~~~~~~~~~~~~~~~~~~~~~~~~~~~
\left.
 +~ G_2^{\pi\pi(\pi NN)}\!(x_{ij})
T_{ij} \right]\tau^+_i\tau^+_j  | \Psi_{A,Z} >.
\end{eqnarray}

We can compare the rates of \Eq{zeroderivativerate} and
\Eq{twoderivativerate} by assuming that all dimensionless constants
are of the order of unity with $1/\rho_{ij}\sim{m_\pi}$ and
${\lh\sim 1}$ GeV, and that the nuclear matrix elements cancel in the
ratio:
\begin{eqnarray}
\frac{\text{\Eq{zeroderivativerate}}}{\text{\Eq{twoderivativerate}}} \sim
\frac{\Lambda_{\text{H}}^4 }{
 m_\pi^4  }\approx
10^3.
\end{eqnarray}
Note that this ratio agrees with our expectation based on power counting.
We end this subsection by emphasizing that \Eq{twoderivativerate} is
not the general formula for the $\nbb$-decay half-life at NNLO (which
must include all contributing terms including loops, recoil effects,
$NNNNee$ terms and $m_\pi^2$ corrections) since
the LO contributions should be added if they do not vanish from symmetry
considerations before squaring the amplitude.

\section{Particle physics models}\label{sec4}

While our discussion so far has been quite general and independent of
the underlying physics of the lepton-number violation,  we apply in
this section our EFT analysis to
two particle physics models:
RPV SUSY and the left-right symmetric (LRS) model.

\subsection{RPV SUSY}

R-parity-violating supersymmetry can contribute to $\nbb$-decay
through diagrams like the one in \Fig{intrographs}b.  Since
supersymmetric particles are heavy, their internal lines can be shrunk
to a point in tree level diagrams yielding operators that involve
only quarks and leptons.    When the RPV
superpotential is expanded to yield a lepton number violating
Lagrangian, and a Fierz transformation is used to separate leptonic
from quark currents, the result is \cite{Faessler:1996ph}
\begin{eqnarray}\label{rpvlag}
{\cal{L}}_{qe}=\frac{\gf^2}{2M}\bar{e}(1+\gamma^5)e^c
\left[
(\eta_{\tilde{q}} + \eta_{\tilde{f}})(J_PJ_P +J_SJ_S)
-\frac{1}{4}\eta_{\tilde{q}}J_T^{\mu\nu}J_{T\mu\nu}
\right],
\end{eqnarray}
where
\begin{eqnarray}
J_P=\bar{q}\gamma^5\tau^+ q,~~ J_S=\bar{q}\tau^+ q,  ~~
J_{T}^{\mu\nu}=\bar{q}\sigma^{\mu\nu}(1+\gamma^5)\tau^+ q,
\end{eqnarray}
and $\eta_{\tilde{q}},\eta_{\tilde{f}}$ are quadratic functions
of the RPV SUSY parameter, $\lambda^\prime_{111}$ defined in
Ref.~\cite{Faessler:1996ph}:
\begin{eqnarray}
\label{eq:etadef}
\eta_{\tilde{k}} & = & \frac{2\pi}{ 9} \frac{|\lambda^{\prime}_{111}|^2M}{
G_F^2m_{\tilde{q}}^4}\left[2\alpha_s \frac{1}{
m_{\tilde{g}}}+\cdots\right],~~~\text{with}~\tilde{k}=\tilde{q},\tilde{f}.
\end{eqnarray}
Here $M$ is the nucleon mass, $m_{\tilde{q}}$ is a first generation squark
mass,  $m_{\tilde{g}}$ is the
gluino mass, $\alpha_s$ is the running SU(3)$_C$ coupling, and the $+\cdots
$ indicate contributions
involving the first generation sleptons and lightest
neutralino\footnote{The slepton/neutralino terms -- which have 
  complicated expressions -- cause
  $\eta_{\tilde{q}}\neq\eta_{\tilde{f}}$.  We have only shown the
  gluino contributions for illustrative purposes.}. Note that
the dependence on $G_F$
and $M$ cancel from Eq.~(\ref{rpvlag}), so that the effective lepton-quark
$0\nu\beta\beta$-decay operator depends on five inverse powers of SUSY
masses.

It is useful to  rewrite \Eq{rpvlag} in terms of our operators
${\cal{O}}_{i\pm}^{++}$:
\begin{eqnarray}\label{Orpvlag}
{\cal{L}}_{qe}=\frac{\gf^2}{2M}\bar{e}(1+\gamma^5)e^c
\left[
\frac{1}{2}(\eta_{\tilde{q}} + \eta_{\tilde{f}})
{\cal{O}}_{2+}^{++}
-\frac{3}{14}\eta_{\tilde{q}}
\left(
{\cal{O}}_{2+}^{++} - {\cal{O}}_{2-}^{++}
\right)
\right].
\end{eqnarray}
The first thing to note is that ${\cal{O}}_{2-}^{++}$ can be neglected
for $0^+\to 0^+$ nuclear transitions.
Secondly, from Table~\ref{qqeehheetable} we see that
${\cal{O}}_{2+}^{++}$ gives rise
to LO $\pi\pi ee$ and NLO $NN\pi ee$ operators and therefore
contributes to the long range $\nbb$-decay operator of
Fig.~\ref{longrangediag}a that is enhanced
relative to the short range interaction
of \Fig{longrangediag}d as observed
by direct calculation in Ref.~{\cite{Faessler:1996ph}, but derived
with different assumptions about the scaling of the LEC.

From Eqs.~(\ref{pipiquarklag}), (\ref{loppeelag}) and (\ref{dimloppeelag}),
it follows that the LO $\pi\pi ee$ operator contributes dominantly to
the $\nbb$-decay in RPV SUSY.  The corresponding half-life formula is
\Eq{zeroderivativerate} with $\beta_1=\beta_2$ and with the
substitution
\begin{eqnarray}
\frac{1}{\lbb}\to \frac{1}{4M}
\left(\frac{4}{7}\eta_{\tilde{q}}+\eta_{\tilde{f}}\right).
\end{eqnarray}
Obviously, a lower limit on the half-life can be interpreted as an
upper limit on the coupling constants $\eta_{\tilde{q}}$ and
$\eta_{\tilde{f}}$.  Making further assumptions about masses of SUSY
particles, one can ultimately obtain model-dependent upper limits on
the coupling constant $\lambda^\prime_{111}$ as discussed in
Ref.~\cite{Faessler:1998qv}.

Next, let us compare the scaling rules used here and in
Refs.~\cite{Faessler:1996ph} and \cite{Faessler:1998qv}.  In the
previous section, we used NDA to
extract the relevant scales out of the dimensionful LEC's by using the
scaling rule \Eq{georgi}.  The alternative method used in
Ref.~\cite{Faessler:1996ph} was to calculate the quark operator
matrix element in the vacuum insertion approximation (VIA) and match
the result to the hadron operator matrix element.

Specifically, for the LO $\pi\pi ee$ operator of \Eq{loppeelag} we
found that the dimensionful LEC scaled as $\lh^2f_\pi^2$ while the VIA would
predict\footnote{Note that we do not take into account the color
factor 8/3 of Ref.~\cite{Faessler:1996ph} since it is a number of
${\cal{O}}(1)$ which does not involve any mass scale.  It can
therefore be absorbed in the LEC's which are undetermined.  See also
the footnote below \Eq{LOquarkop5}.}:
\begin{eqnarray}
\text{LEC's}\sim \langle\pi^+|J_PJ_P|\pi^-\rangle &\approx&
\langle\pi^+|J_P|0\rangle\langle0|J_P|\pi^-\rangle
\nonumber
\\
&=& -2f_\pi^2 \frac{m_\pi^4}{(m_u+m_d)^2}~~,
\end{eqnarray}
where $m_{u,d}$ are the light quark masses.  Taking
$\lh=\Lambda_\chi=4\pi f_\pi$, the chiral
symmetry breaking scale,  and $m_u+m_d=11.6$ MeV we find
\begin{equation}\label{lambdahratio}
\frac{\text{NDA}}{\text{VIA}} \sim
\frac{(4\pi f_\pi)^2f_\pi^2}{ 2f_\pi^2 \frac{m_\pi^4}{(m_u+m_d)^2} }=0.7~.
\end{equation}
The NDA scaling is thus slightly smaller than that obtained from the
VIA.  Although they give results of the same order, VIA has proved to
be unreliable in other contexts (see, {\em e.g.}, the study of
rare kaon decays in Ref.~\cite{Donoghue:dd}) .
We will therefore use NDA in what follows.

Referring to Table~\ref{qqeehheetable}, it follows that there should
be  additional, subdominant
contributions from the operator $\pi\pi ee$ and from the
$NN\pi ee$ operator at NNLO.  The NNLO
contributions from the $NN\pi ee$ vertex were considered in
Ref.~\cite{Faessler:1998qv} where  detailed numerical evaluations
showed that they contribute on average about thirty  times
less then the LO contribution.  Our systematic analysis leads to the
same qualitative conclusion (namely with regards to the NNLO
suppression of $p^2/\lh^2$ with respect to the LO), but differs from
Ref.~\cite{Faessler:1998qv} in some respects.

First of all, not all NNLO contributions were included.  In
particular, as pointed out above, the NNLO $\pi\pi ee$ operator
contributes to $\nbb$-decay at the same order as the $NN\pi ee$
operator (called 1$\pi$ in Ref.~\cite{Faessler:1998qv}) and the
form-factors $G_{1,2}^{\pi\pi}$ should be included.

Secondly, our analysis shows that the
$NNNNee$ operator (the only one considered previously in this type of
analysis) gives contributions at NNLO.\footnote{We note that the
  long-range operators considered in Ref.~\cite{Simkovic:1999re}
  through the induced pseudoscalar coupling terms of the nucleon
  current correspond to the NNLO contributions of \Eq{nnloppeelag}.
  The results presented by the authors of Ref.~\cite{Simkovic:1999re}
  in the form-factor approach are compatible with the EFT analysis
  given here since they only considered left-handed hadronic
  currents.}
In Refs.~\cite{Faessler:1996ph} 
the suppression of that operator relative to the LO $\pi\pi ee$
contribution was only by a factor of ten for $^{76}$Ge which is
larger than what would be expected from our power counting (see also
Ref.~\cite{Wodecki:1999gu}).  However, 
this suppression is still in qualitative agreement with our analysis
keeping in mind that considerable uncertainty remains
in the evaluation of nuclear matrix elements.  Furthermore,
although the traditional
method of calculating the short-range $\nbb$-decay
operator using dipole form-factors~\cite{Vergados:1981bm} may yield
results of the correct order, the method is unsystematic with
uncontrollable errors that cannot be easily estimated.

\subsection{Left-right symmetric model}

We consider LRS models that contain a heavy
right-handed neutrino,  and  mixing between the right-handed and
left-handed gauge bosons with $g_{\text{L}}\approx g_{\text{R}} = g$ where
$g_{\text{L}}$ and $g_{\text{R}}$ are the left-handed and right-handed
gauge couplings.  The LRS Lagrangian is taken to be invariant under
$\text{SU(2)}_{\text{L}}\times\text{SU(2)}_{\text{R}}\times
\text{U(1)}_{B-L}$ where $B,L$ are the baryon, lepton numbers
respectively.
We will not be concerned with the CP-violating phases of the mixing
matrix $U^{\text{R}}$ of the right-handed quark generations (the
right-handed equivalent of the Cabbibo-Kobayashi-Maskawa matrix,
denoted here $U^{\text{L}}$) nor
the precise nature of the relationship between $U^{\text{R}}$ and
$U^{\text{L}}$ ({\it e.g.}, manifest versus pseudo-manifest LRS model)
as the order of magnitude of the constraints obtained from experiments
are broadly robust to the different possibilities
\cite{Beg:ti,Abachi:1995yi,Abe:1994zg,Langacker:1989xa}. We will use
the standard Higgs sector composed of a
left-handed triplet, $\Delta_{\text{L}}$, a right-handed
triplet, $\Delta_{\text{R}}$, and a multiplet, $\Phi$, that respectively
transform under
$\text{SU(2)}_{\text{L}}\times\text{SU(2)}_{\text{R}}\times
\text{U(1)}_{B-L}$ according to $(L,R,Y)$= (3,1,2), (1,3,2) and
(2,2,0).  Their vacuum expectation values are
\begin{eqnarray}
\langle\Delta_{\text{L}}\rangle =
\left(
\begin{array}{c}
0
\\
0
\\
\Delta_{\text{L}}^0
\end{array}
\right),~~~~~~
\langle\Delta_{\text{R}}\rangle =
\left(
\begin{array}{c}
0
\\
0
\\
\Delta_{\text{R}}^0
\end{array}
\right),~~~~~~
\langle\Phi\rangle =
\left(
\begin{array}{cc}
\kappa & 0
\\
0 & \kappa^\prime
\end{array}
\right).
\end{eqnarray}
Assume the following relation between the gauge and the
mass eigenstates (ignoring the possibility of a CP-violating phase)
\begin{eqnarray}
W_{\text{L}} &=& \cos \zeta  W_1 + \sin\zeta  W_2
\nonumber
\\
W_{\text{R}} &=& -\sin\zeta  W_1 + \cos \zeta  W_2~,
\end{eqnarray}
where $\zeta $ is a small mixing angle between the mass eigenstates
and,
\begin{eqnarray}\label{masszetaexps}
M_{W_1}^2 &\cong& \frac{g^2}{2}
\left(
\kappa^2 + {\kappa^\prime}^2
\right),
\\
M_{W_2}^2 &\cong& \frac{g^2}{2}
\left(
\kappa^2 + {\kappa^\prime}^2 + 2{\Delta^0_{\text{R}}}^2
\right),
\\
\zeta &\cong& \frac{\kappa\kappa^\prime}{{\Delta^0_{\text{R}}}^2}~,
\end{eqnarray}
where  $M_{W_{1,2}}$  are the masses of $W_{1,2}$.  From these
equations and the fact that
$|\kappa^2~+~{\kappa^\prime}^2|/2\ge~|\kappa\kappa^\prime|$, we
immediately obtain 
the important relation first derived in
Ref.~\cite{Masso:1984bt},\footnote{From here on, $\zeta$ will
  exclusively denote the magnitude of the mixing angle.}
\begin{eqnarray}\label{masso}
\lambda \equiv \left(\frac{M_{W_1}}
{M_{W_2}}\right)^2 \ge \zeta~.
\end{eqnarray}
Turning to experimental bounds on the masses and mixing angles, we
will use for the lower limit on the right handed gauge boson
$M_{W_{2}}> 715$~GeV \cite{Hagiwara:fs}, which corresponds roughly to
\begin{eqnarray}\label{mzetalimits}
\lambda < 10^{-2},
\end{eqnarray}
To put limits on the mixing angle, we use recent
results from superallowed $0^+\to 0^+$ $\beta$-decay in
Ref.~\cite{Hardy:ci} that imply a violation of the unitarity of the
CKM matrix at the 95\% confidence level.  In the LRS model, unitarity
can be restored by taking a positive value for the mixing angle with
magnitude
\begin{equation}\label{unitarity}
\zeta  =0.0016\pm 0.0007~,
\end{equation}
given that one has
\begin{equation}
|V_{ud}|^2+|V_{us}|^2+|V_{ub}|^2=0.9968\pm 0.0014~,
\end{equation}
in the Standard Model only~\cite{Hardy:ci} .
A range of $2\times 10^{-4}\le \zeta \le 3\times 10^{-3}$ is allowed
at 95\% confidence level.  Note that the discrepancy in the unitarity
condition cannot be resolved by adjusting $\lambda$ because
it enters the ordinary $\beta$-decay amplitude quadratically and, thus,
produces a correction smaller than $10^{-4}$ [see \Eq{mzetalimits}].
In what follows, we will consider the range $0\le\zeta\le 3\times
10^{-3}$ and use the central value of
\Eq{unitarity} for some specific estimates.  Note that for the
central value of $\zeta$ of \Eq{unitarity},
we obtain an upper limit on $M_{W_2}$ from \Eq{masso} of
\begin{eqnarray}\label{uplimmw2}
M_{W_2} \le  M_{W_1}/\sqrt{\zeta} \rightarrow M_{W_2} \le 2~\text{TeV,~
 for}~ \zeta=0.0016~.
\end{eqnarray}
With these bounds on
$M_{W_{2}}$ and $\zeta$, we can now estimate the relative order of
magnitude of the graphs of \Fig{lrsgraphs}.

When the right-handed neutrino and $W_{L,R}$ are integrated out,
the amplitude of  \Fig{lrsgraphs}a
reduces to an operator of the form
${\cal{O}}_{3+}^{++}$ while \Fig{lrsgraphs}b reduces to an operator of
the form ${\cal{O}}_{1+}^{++}$.\footnote{Recall that the parity-odd
  $LL/RR$   operator ${\cal{O}}_{3-}^{++}$ is suppressed at NNLO.}  In
previous treatments of $\nbb$-decay, only
graph~\ref{lrsgraphs}a with right-handed interacting currents is
considered and the impact of $W_L$-$W_R$ mixing is neglected.
Our analysis of the previous sections
implies that the hadronic operators generated by
${\cal{O}}_{3+}^{++}$
are suppressed by a factor of $p^2/\lh^2\sim 10^{-2}$ relative
to those generated by ${\cal{O}}_{1+}^{++}$.  Hence, taking into
account the fact that the coupling of a (right)left-handed current with a
($W_1$)$W_2$ involves a suppression factor of $\zeta $ while a
$W_2$ internal line involves a suppression factor of $\lambda$, we
expect the $\pi\pi$ operators generated by these quark operators to
scale as
\begin{eqnarray}\label{scalefac}
M_{\ref{lrsgraphs}\text{a)}}^{\text{(LL)}} \sim \zeta
^2\frac{p^2}{\lh^2}< 10^{-8}  ,
~~~~~& &
M_{\ref{lrsgraphs}\text{a)}}^{\text{(RR)}} \sim \lambda^2\frac{p^2}{\lh^2}<
10^{-6}  , ~~~~~
\nonumber
\\
M_{\ref{lrsgraphs}\text{b)}}^{\text{(LR)}} &\sim& \lambda \zeta  < 10^{-5}  ,
\end{eqnarray}
with all else assumed equal.  Therefore, even if $\zeta $ is ten times
smaller than the central value in Eq.~(\ref{unitarity}), the contribution
stemming from the
mixing of left-handed and right-handed gauge bosons is still
non-negligible.  It may even be dominant.

\begin{center}
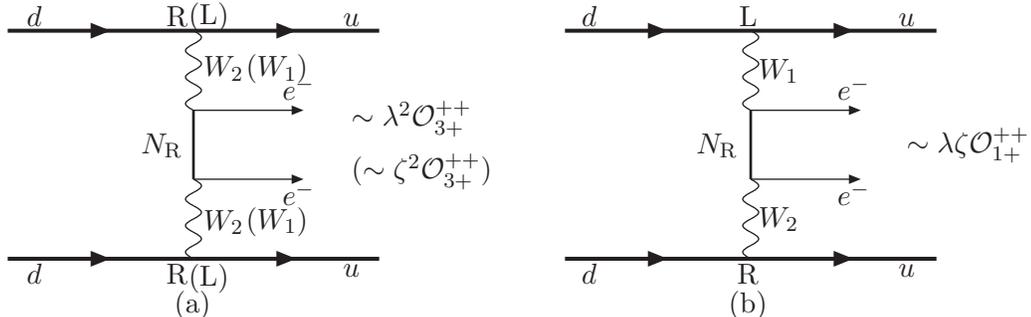
\begin{figure}\caption{ \sl \footnotesize Left-right symmetric model
    graphs.  \Fig{lrsgraphs}a involves the interaction of two
    right-handed (left-handed) currents while \Fig{lrsgraphs}b
    depicts the interaction of left-handed and right-handed currents.
    }\label{lrsgraphs}
\begin{picture}(360,100)(0,0)
%
%
{\SetWidth{1.5}
\ArrowLine(0,90)(70,90)
\ArrowLine(70,90)(140,90)
\ArrowLine(0,4)(70,4)
\ArrowLine(70,4)(140,4)}
\Photon(70,90)(70,60){3}{3}
\Photon(70,4)(70,34){3}{3}
{\SetWidth{1} \Line(70,60)(70,34)}
\LongArrow(70,60)(110,60)
\LongArrow(70,34)(110,34)
\Text(130,57)[l]{$\sim  \lambda^2 {\cal{O}}_{3+}^{++}$}
\Text(130,37)[l]{($\sim  \zeta ^2 {\cal{O}}_{3+}^{++}$)}
\Text(74,76)[l]{$W_2$}
\Text(74,18)[l]{$W_2$}
\Text(90,76)[l]{$(W_1)$}
\Text(90,18)[l]{$(W_1)$}
\Text(64,92)[b]{R}
\Text(64,2)[t]{R}
\Text(76,90)[b]{(L)}
\Text(76,2)[t]{(L)}
\Text(70,-8)[t]{(a)}
\Text(10,92)[b]{$d$}
\Text(10,2)[t]{$d$}
\Text(130,92)[b]{$u$}
\Text(130,2)[t]{$u$}
\Text(110,64)[b]{$e^-$}
\Text(110,33)[t]{$e^-$}
\Text(66,47)[r]{$N_{\text{R}}$}
%
%
{\SetWidth{1.5}
\ArrowLine(210,90)(280,90)
\ArrowLine(280,90)(350,90)
\ArrowLine(210,4)(280,4)
\ArrowLine(280,4)(350,4)}
\Photon(280,90)(280,60){3}{3}
\Photon(280,4)(280,34){3}{3}
{\SetWidth{1} \Line(280,60)(280,34)}
\LongArrow(280,60)(320,60)
\LongArrow(280,34)(320,34)
\Text(340,47)[l]{$ \sim \lambda \zeta  {\cal{O}}_{1+}^{++}$}
\Text(284,75)[l]{$W_1$}
\Text(284,19)[l]{$W_2$}
\Text(280,92)[b]{L}
\Text(280,2)[t]{R}
\Text(280,-8)[t]{(b)}
\Text(220,92)[b]{$d$}
\Text(220,2)[t]{$d$}
\Text(340,92)[b]{$u$}
\Text(340,2)[t]{$u$}
\Text(320,64)[b]{$e^-$}
\Text(320,33)[t]{$e^-$}
\Text(276,47)[r]{$N_{\text{R}}$}
\end{picture}
\end{figure}
\end{center}
\begin{center}
\begin{figure}\caption{ \sl \footnotesize Constraints on the
    right-handed weak boson and neutrino masses (in TeV) in the LRS
    model.  The
    solid lines stem from the vacuum stability (V.S.) constraint of
    \Eq{vacstab} while the hyphenated lines correspond to limits
    imposed from $\nbb$-decay and \Eq{masso} with the following
    values of mixing angle from longest to shortest dashes:
    $\zeta_i=\{3\times 10^{-3},1.6\times 10^{-3},0\}$ with
    $i=1,2,3$.  Graphs (a), and (b)
    correspond to cases 1 and 2 of the text,
    respectively.  Note that the value of the mixing
    angle $\zeta_3=0$ cannot occur for case 2 without simultaneously
    taking $M_{W_2}$ to infinity, while $\zeta_2$
    corresponds to the central value obtained from CKM
    unitarity.  The arrows indicate the lower bound  $M_{W_2}\ge$715 GeV
    imposed by direct searches.  The shaded, triangular regions in the
    graphs are the allowed 
    values of the masses if the mixing angle is $\zeta_2$.}\label{WrvsNr1}
\begin{picture}(460,150)(0,0)
%
%
\scalebox{0.8}[0.8]{\includegraphics{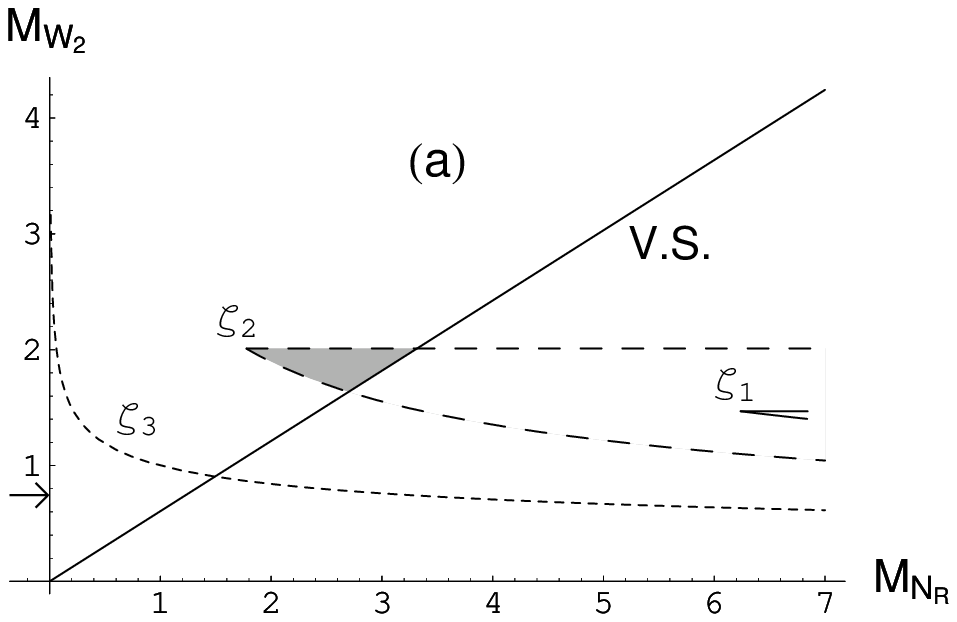}}
\scalebox{0.8}[0.8]{\includegraphics{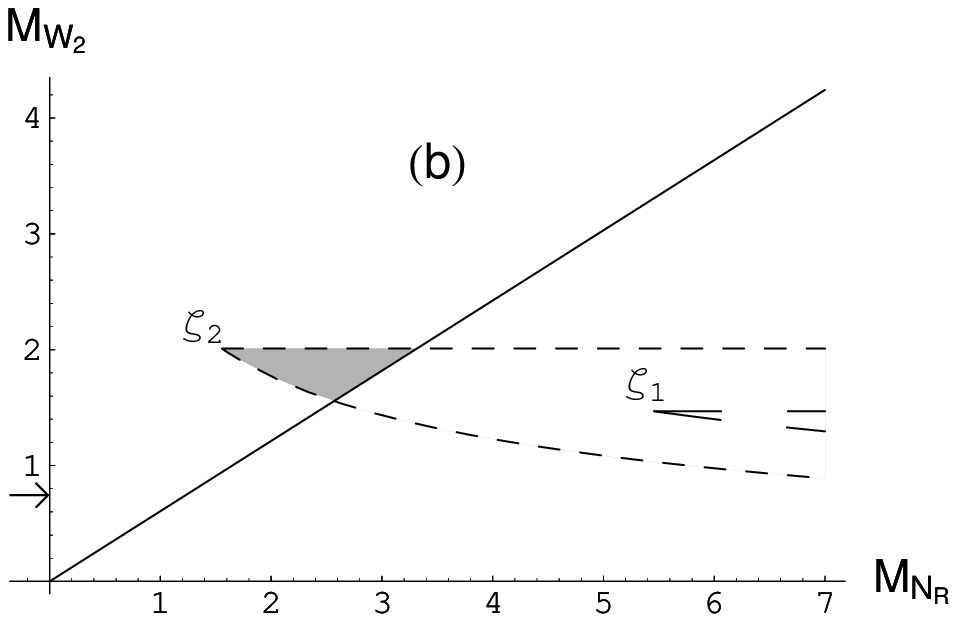}}
%
\end{picture}
\end{figure}
\end{center}

 Such  analysis may modify two constraints that relate the
right-handed weak boson and neutrino masses, $M_{W_{\text{2}}}$ and
$M_{N_{\text{R}}}$ respectively\footnote{For illustrative
  purposes, we assume the existence
  of only one right-handed neutrino.}.

The first constraint stems from the requirement that the vacuum
expectation value of the
Higgs field $\Delta_{\text{R}}$ be a true minimum of the Higgs potential that
generates the
masses of the right-handed particles~\cite{Mohapatra:pj}.  The vacuum
is then stable against collapse.  This imposes stringent
constraints on the one-loop corrections to the effective
potential~\cite{Coleman:jx,Weinberg:pe,Hung:1979dn}.  In particular,
the loop corrections will involve terms of the form $k \Delta_{\text{R}}^4
\ln (\Delta_{\text{R}}^2/{\Delta_{\text{R}}^0}^2)$ where 
$k$ is a constant that depends on the particle masses.  For
the vacuum to be stable at large values of $\Delta_{\text{R}}$, $k$ must be
positive to  ensure that the minimum at the VEV is a true
minimum and not simply a local minimum.  The condition $k >0$ is
equivalent to a 
condition on the masses.  Following this formalism allows us to derive
a relationship between  $M_{W_2}$ and $M_{N_{\text{R}}}$:
\begin{eqnarray}\label{vacstab}
1.65M_{W_{\text{2}}}\ge M_{N_{\text{R}}}.
\end{eqnarray}
This constraint is represented in the graphs of \Fig{WrvsNr1} by the
fact that no value of $(M_{N_{\text{R}}},M_{W_{\text{2}}})$ below the
solid lines is allowed\footnote{In Ref.~\cite{Mohapatra:pj}, the
  constraint that appears is $0.95M_{W_{\text{2}}}\ge
  M_{N_{\text{R}}}$, the result of a typo~\cite{private}.}.

A second relationship constraining $M_{W_{\text{2}}}$ and
$M_{N_{\text{R}}}$  in the LRS model
with mixing can be inferred from experimental limits on
$\nbb$-decay~\cite{Mohapatra:pj,Hirsch:1996qw} from
\Eq{zeroderivativerate} with $\lbb=M_{N_{\text{R}}}$ and
choosing $\beta_1=\beta_2=1$
\begin{eqnarray}\label{nbbconstraint}
|\zeta\lambda\pm\delta(\lambda^2+\zeta^2)|^2
< \frac{9}{2}\frac{M_{N_{\text{R}}}^2}{\lh^4 G_{0\nu}^{(A,Z)}
 |{\cal{M}}_0^{(A,Z)}|^2 T_{1/2}^{(A,Z)}} \equiv {\nu^{(A,Z)}}^2,
~~~~~~~~~~~~~~& &
\\
G_{0\nu}^{(A,Z)}=(\gf \cos\theta_{\text{C}}g_{\text{A}})^4
\left(\frac{\hbar c}{R}\right)^2\frac{1}{32\pi^5\hbar\ln 2}
\times
~~~~~~~~~~~~~~~~~~~~~~~& &
\nonumber
\\
     \int_{m_e}^{E_{\beta\beta} - m_e}\text{d}\!E_1
  F\!(Z+2,E_1)F\!(Z+2,E_2)p_1E_1p_2E_2~,
& &
\end{eqnarray}
where $\nu^{(A,Z)}$ is defined by \Eq{nbbconstraint}, $\lh\approx 1$
GeV, $T_{1/2}^{(A,Z)}$ is the current limit on the
half-life of the $\nbb$-decay transition of a
nucleus $(A,Z)$ and where the functions $G_{0\nu}^{(A,Z)}$ were
tabulated in Ref.~\cite{Doi:1985dx} for various nuclei.
The matrix element ${\cal{M}}_0^{(A,Z)}$ is defined
in \Eq{matrixelement}.\footnote{From
 here on, we take $\cos\theta_{\text{C}}=1$.}  In
\Eq{nbbconstraint} we have made explicit the scaling factors of
\Eq{scalefac} and also introduced a factor $\delta$ which
parametrizes the $p^2/\lh^2$ suppression of the NNLO $\nbb$-decay
operators
relative to the LO operators.  As mentioned above, the numerical
evaluations in Ref.~\cite{Faessler:1998qv} suggest that $\delta
\approx 1/30$ which is the conservative number we will use.  Thus, the
$\lambda^2$ term stems from the exchange of two
$W_2$'s  while the $\zeta^2$ term comes from the
exchange of two $W_1$'s where $\zeta$, being the magnitude
of the mixing angle, is always positive.
The relative sign between the $\zeta\lambda$ and $\delta(\lambda^2+\zeta^2)$
terms on the LHS of
\Eq{nbbconstraint} cannot be predicted by EFT since we do not know the
sign of the LEC's.

For the values of half-life, $G_{0\nu}^{(A,Z)}$ and
${\cal{M}}_0^{(A,Z)}$, we will use the ones determined for
$^{76}$Ge
\begin{eqnarray}\label{germanium}
T_{1/2}^{\text{Ge}}\ge 1.9 \times 10^{25}\text{yrs},~~~~~
(G_{0\nu}^{\text{Ge}})^{-1}=4.09\times
  10^{25}\text{eV}^2~\text{yrs},~~~~~
{\cal{M}}_0^{\text{Ge}}=2,
\end{eqnarray}
where we extracted the value of ${\cal{M}}_0^{\text{Ge}}$ from the
value of ${\cal{M}}^{2\pi}$ calculated in Ref.~\cite{Faessler:1998qv}
and the limit on the half-life is at 90\% confidence
level~\cite{Klapdor-Kleingrothaus:2000sn}.  With these numbers,
\Eq{nbbconstraint} becomes:
\begin{eqnarray}\label{nbbconstraint2}
|\zeta\lambda \pm \delta(\lambda^2+\zeta^2)|
< \nu^{\text{Ge}} =
\sqrt{\frac{9}{3.8}}\left(\frac{M_{N_{\text{R}}}}{\text{TeV}}\right)
10^{-6},
\end{eqnarray}
In the limit $\zeta\to 0$ we obtain
\begin{eqnarray}\label{mwrconstraint}
M_{W_{\text{2}}}>
 \left(\delta\sqrt{\frac{3.8}{9}}
\frac{\text{TeV}}{M_{N_{\text{R}}}}10^{6}\right)^{1/4} M_{W_{\text{1}}}
\cong\left(\frac{\text{TeV}}{M_{N_{\text{R}}}}\right)^{1/4}\text{TeV}.
\end{eqnarray}
Our result is slightly smaller then the result obtained in
Refs.~\cite{Mohapatra:pj,Hirsch:1996qw} for zero mixing angle.   In
Refs~\cite{Mohapatra:pj} this constraint  was
calculated with the short range  NNLO $NNNNee$ operator of
\Fig{longrangediag}d using the dipole form factor approach.  Note
that we can reproduce exactly the values given in
Refs.~\cite{Mohapatra:pj,Hirsch:1996qw} by slightly adjusting the
unknown constants $\beta_1,\beta_2$ in \Eq{zeroderivativerate}.

To extract the constraint imposed by \Eq{nbbconstraint2} on
$M_{N_{\text{R}}}$ and $M_{W_{\text{2}}}$, we need to consider three
cases:
\begin{itemize}

\item (1) the LO and NNLO terms have the same sign which corresponds
to taking the plus sign in \Eq{nbbconstraint2},

\item (2)  they have opposite
signs with $\zeta\lambda > \delta(\lambda^2+\zeta^2)$, and

\item (3)  they have
opposite signs with $\zeta\lambda < \delta(\lambda^2+\zeta^2)$.
\end{itemize}
We note that in all three cases, the upper limit on $M_{W_2}$ for
$\zeta > 0 $ implied by \Eq{masso} always holds.

{\it Case 1:} When solving the quadratic equation in $\lambda$, we must
keep the root that has the same limit as \Eq{nbbconstraint2} when
$\delta,\zeta\to 0$,
\begin{eqnarray}\label{beta1}
\zeta\le \lambda \le \frac{1}{2\delta}
  \left(-\zeta+\sqrt{(1-4\delta^2)\zeta^2 +
  4\delta \nu}\right),
\end{eqnarray}
where we used \Eq{masso} to obtain the first inequality.  The first thing
to note is that Eqs. (\ref{nbbconstraint2}-\ref{beta1})
impose a lower-limit on the mass of the right-handed neutrino
\begin{eqnarray}\label{lowerlim}
\frac{M_{N_{\text{R}}}}{\text{TeV}} > \sqrt{\frac{3.8}{9}} 10^6
(1+2\delta) \zeta^2 \cong 1.8~,
\end{eqnarray}
assuming the central value of \Eq{unitarity}.
This lower limit only
depends on the mixing angle since $\delta$ can
in principle be calculated.  In \Fig{WrvsNr1}a, the constraint
\Eq{beta1} is plotted for three values of the mixing angle
$\zeta_i=\{3\times 10^{-3},1.6\times 10^{-3},0\}$.

In \Fig{WrvsNr1}a, we see that the larger the mixing
angle, the larger the parameter space that is ruled out.
In particular, for $\zeta_1$, the largest angle that we are
considering, the region allowed by Eqs.~(\ref{beta1})
and~(\ref{lowerlim})  is located below the constraint imposed by
vacuum stability.  Hence, a value of the mixing angle as large as
$\zeta_1$ is excluded.  In contrast, the central mixing angle value
from CKM unitarity, $\zeta_2$, allows for a triangular region [bordered
by the vacuum stability curve and Eqs.~(\ref{beta1})] of possible values
for the masses.  In particular, for $\zeta_2$, we note that not only
do we have the upper-limit of \Eq{uplimmw2}, but we also have
$M_{W_2}\ge 1.6$~TeV and $M_{N_{\text{R}}}\le 3.2$~TeV, which would
constitute  more stringent limits than that obtained from direct
searches so far.
For zero mixing angle, the entire region that is
simultaneously above the vacuum stability curve and the curve
stemming from \Eq{mwrconstraint} is allowed.
Thus, in general, as the mixing angle
increases, the allowed region of parameter space shrinks while
the minimum value of $M_{W_{\text{2}}}$ increases.  The
maximum mixing angle that results in a non-vanishing allowed
region\footnote{Actually, a point in this case.}
\begin{eqnarray}\label{maxangle1}
\zeta \le 2.2\times 10^{-3},
~~~\text{with}~M_{W_{\text{2}}}\cong 1.7~\text{TeV},~~
M_{N_{\text{R}}}\cong 2.8~\text{TeV}.
 \end{eqnarray}

{\it Case 2:} The condition of validity for this case, $\zeta\lambda >
\delta(\lambda^2+\zeta^2)$, rules out the positive root of the
quadratic equation in $\lambda$,
\Eq{nbbconstraint2}.  The limits on $\lambda$ are then
\begin{eqnarray}\label{beta2}
\zeta \le \lambda \le
  \frac{1}{2\delta} \left(\zeta-\sqrt{(1-4\delta^2)\zeta^2 -
  4\delta \nu}\right).
\end{eqnarray}
We note that \Eq{beta2} imposes upper and lower limits on both
$M_{N_{\text{R}}}$ and $M_{W_{\text{2}}}$,
\begin{eqnarray}\label{upperlim}
\sqrt{\frac{3.8}{9}} 10^6
(1-2\delta) \zeta^2 \le
\frac{M_{N_{\text{R}}}}{\text{TeV}} \le
\sqrt{\frac{3.8}{9}} 10^6 \frac{1}{4\delta}(1-4\delta^2) \zeta^2 ,
~~~~~
\sqrt{\frac{2\delta M_{W_1}^2}{\zeta}}  \le
M_{W_{\text{2}}}~.
\end{eqnarray}
For $\zeta_2$, we obtain in particular, 1.6~TeV$\le
M_{N_{\text{R}}}\le 12$~TeV and $M_{W_{\text{2}}}\ge 0.51$~TeV.  Note
that the upper limit on $M_{N_{\text{R}}}$ for $\zeta_2$ is well above
the constraint stemming from vacuum
stability, \Eq{vacstab}, combined with the upper limit on $M_{W_2}$ given
in \Eq{uplimmw2}.  Eqs.~(\ref{upperlim}) also implies a new
relationship between $M_{N_{\text{R}}}$ and $M_{W_{\text{2}}}$
applicable only to case 2,
\begin{eqnarray}\label{newmnrmw2}
M_{W_{\text{2}}}\le
\left(
\sqrt{\frac{3.8}{9}} \frac{10^6 \text{TeV}}{4\delta
  M_{N_{\text{R}}}}
\right)^{\frac{1}{4}}
M_{W_1} \cong 3.8\left(
\frac{\text{TeV}}{M_{N_{\text{R}}}}
\right)^{\frac{1}{4}}\text{TeV},
\end{eqnarray}
where we neglected the $4\delta^2$ term.

From the plot in \Fig{WrvsNr1}b, the same analysis as in case 1
follows: as the mixing angle increases, the region of allowed values
for the masses shrinks.  As in case 1, $\zeta_1$ is already excluded
while $\zeta_2$ allows for a triangular region of possible values for
the masses.  We note that \Eq{newmnrmw2} does not further constrain
the allowed region of parameter space and has been included here for
completeness.   For this case, the
maximum mixing angle is calculated to be,
\begin{eqnarray}\label{maxangle2}
\zeta \le 2.1\times 10^{-3},
~~~\text{with}~M_{W_{\text{2}}} \cong 1.8~\text{TeV},~~
M_{N_{\text{R}}} \cong 2.9~\text{TeV},
 \end{eqnarray}
which are similar to the values found for case 1.

{\it Case 3:} For the case $\zeta\lambda < \delta(\lambda^2+\zeta^2)$,
we must keep the root that gives the correct upper-limit when $\zeta\to 0$
since now the limit $\delta \to 0$ cannot be taken.  With the
constraint on $\lambda$ stemming from
the condition of validity of this case, $\zeta\lambda <
\delta(\lambda^2+\zeta^2)$, the inequalities satisfied by $\lambda$ are
\begin{eqnarray}\label{beta3}
\lambda \le \frac{\zeta}{2\delta}\left(1-\sqrt{1-4\delta^2}\right),~~~
\frac{\zeta}{2\delta}\left(1+\sqrt{1-4\delta^2}\right) \le
\lambda \le \frac{1}{2\delta} \left(\zeta+\sqrt{(1-4\delta^2)\zeta^2 +
  4\delta \nu}\right).
\end{eqnarray}
Thus, values of $\lambda$ located between the roots
$\lambda_{\pm}=\zeta/(2\delta)(1\pm\sqrt{1-4\delta^2})$ are
excluded.\footnote{Since $1/(2\delta)
  \left(\zeta+\sqrt{(1-4\delta^2)\zeta^2 +  4\delta \nu}\right)\gg
  \zeta/(2\delta)\left(1-\sqrt{1-4\delta^2}\right)$  for all non-zero
  values of $\zeta$ and $\nu$, we need only be concerned with the
  $\lambda_-$ upper limit on $\lambda$.}
Note that
for the two non-zero angles considered in  \Fig{WrvsNr1}, the ranges
defined by $\lambda\ge \lambda_+$ have already been ruled out
by direct searches of right-handed
bosons~\cite{Abachi:1995yi} and we are left with the first constraint
of Eqs.~(\ref{beta3}) which does not depend on limits from $\nbb$-decay.
However case three appears to be entirely
ruled out by \Eq{masso}.
Indeed, approximating the remaining constraint of \Eq{beta3} to
$\lambda<\delta\zeta$, we see that both constraints cannot be
satisfied simultaneously.



From \Fig{WrvsNr1} and the three cases considered above, it follows
that the effect of mixing on the mass constraint can be very important
-- a point not recognized previously.  In particular, we see that non-zero
mixing angles will generally exclude much of the parameter space by
imposing much more stringent constraints on the masses and that the
mass of the right-handed neutrino is bounded from below.  We also note
that quite generally, the mixing angle is constrained to
be~$\le 2.2\times 10^{-3}$.


We conclude this section by briefly comparing the left-right symmetric
model and RPV SUSY.  We
observe that although both models can contribute to
${\cal{O}}(p^{-2})$ to the operator of \Fig{longrangediag}a, only RPV
SUSY contributes to Figs.~\ref{longrangediag}b,c to  ${\cal{O}}(p^{-1})$ as
discussed in the previous section.  These
results are summarized in Table~\ref{complrsrpv}.

\begin{table}\caption{Order at which the left-right symmetric models
    with/without mixing and RPV SUSY contribute to the $\nbb$-decay
    operators of \Fig{longrangediag}. } \label{complrsrpv}

\begin{tabular}{|c|c|c|c|}\hline
Models
 & \Fig{longrangediag}a
 & \Fig{longrangediag}b,c
 &  \Fig{longrangediag}d\\
\hline
LRSM $\zeta  = 0$
& $p^{0}$
& $p^{0}$
& $p^{0}$
\\
\hline
LRSM $\zeta \neq 0$
& $p^{-2}$
& $p^{0}$
& $p^{0}$
 \\
\hline
RPV SUSY
& $p^{-2}$
& $p^{-1}$
& $p^{0}$
 \\
\hline
\end{tabular}
\end{table}

\section{Conclusions}\label{sec5}

Neutrinoless double beta-decay will continue to probe \lq\lq new" physics
scenarios that violate lepton number for some time to come. The existence of
such scenarios is intimately related to the nature of the neutrino, namely,
whether or not it is a Majorana particle. If a significant signal for
$0\nu\beta\beta$ decay were to be observed, one would know that the neutrino
is a Majorana particle. However, one would not know whether the rate is
dominated by the exchange of a light Majorana neutrino or by some other
L-violating process that is also responsible for generation of the Majorana
mass. Such L-violating processes could involve mass scales
($\Lambda_{\beta\beta}$) well above the weak scale. Thus, it is important
to study the implications of $0\nu\beta\beta$-decay for such scenarios -- a
task which we have undertaken in the present paper.

In doing so, we have applied the ideas of EFT, which is appropriate in this
case because there is a clear distinction of scales:$\Lambda_{\beta\beta}\gg
\Lambda_H\gg p$.  We wrote down all non-equivalent quark-lepton operators
of dimension nine that contribute to $\nbb$-decay, and showed how to
match them to
hadron-lepton operators by using their transformation properties under
parity and chiral SU(2).  We then organized the
hadron-lepton operators ($\pi\pi ee$, $NN\pi ee$ and $NNNNee$) 
in powers of $p/\lh$ and discussed how the symmetries determine
the type of hadronic operators that can be 
generated by each quark operators.  In particular, we demonstrated that the
hadronic operators generated by the interaction of two left-handed or
two right-handed quark currents are always of NNLO.  We also showed
that EFT can classify particle physics models of $\nbb$-decay in terms
of the hadron-lepton operators they can generate and to what order
these operators
enter.  In particular, we found that left-right symmetric models with
mixing can potentially and considerably modify existing constraints on
the masses of the right-handed particles.  Indeed,  a non-zero
mixing angle  gives far more stringent constraints on the allowed
values of the masses of right-handed particles including a correlation
between the mass of the right-handed neutrino and the
mixing angle.  We also found that a necessary condition for the
existence of a region of allowed values of $M_{W_2}$ and $M_{N_{\text{R}}}$
is~$\zeta\le 2.2\times 10^{-3}$.  For RPV SUSY models, we have also
confirmed the previous conclusion that the dominant contribution stems
from the $\pi\pi ee$
operator which leads to more severe constraints on the corresponding
RPV SUSY parameters than traditionally believed.  More generally, with
this EFT analysis and using Table~\ref{qqeehheetable}, it can be
immediately known what hadron-lepton operators can be generated by any
quark-lepton operators appearing in any particle physics model that
gives rise to $\nbb$-decay, and to what order these hadron-lepton
operators will contribute.
Finally, we note that deriving detailed information about a given scenario
for L-violation will require combining information from a variety of
measurements. As our analysis of the left-right symmetric model shows, using
studies of $0\nu\beta\beta$-decay in conjunction with precision electroweak
measurements ({\em e.g.}, light quark $\beta$-decay) and collider experiments
can more severely constrain the particle physics parameter space than can any
individual probe alone. Undertaking similar analysis for other new physics
scenarios and other probes of L-violation constitutes an interesting
problem for future study.

We thank P. Bedaque, M. Butler, R. Mohapatra, and M. Savage for useful
discussions. P.V. thanks Prof. J. Ho\v{r}ej\v{s}\'{\i} for
his hospitality at the Center for Particle and Nuclear Physics, Charles 
University, Prague, Czech Republic.
This work was supported in part under Department of Energy
contracts DE-FG02-00ER41146, DE-FG03-02ER41215, DE-FG03-88ER40397, and
DE-FG03-00ER41132 and NSF award PHY-0071856.

\begin{appendix}

\section{Equivalent and vanishing quark operators}\label{otherquarkops}

All operators proportional to $\bar{e}^c\gamma_\mu e$ and
$\bar{e}^c\sigma_{\mu\nu}e$ vanish identically by virtue of the
fact that the electron fields are Grassmann variables.  For example:
\begin{eqnarray}
\bar{e}^c\gamma_\mu e &=& i e_\alpha \gamma^0_{\alpha\beta}
\gamma^2_{\beta\sigma} \gamma^\mu_{\sigma\delta}e_\delta \nonumber\\ 
&=& -i e_\delta (\gamma^\mu_{\delta\sigma})^{\text{T}}
\gamma^2_{\sigma\beta} \gamma^0_{\beta\alpha} e_\alpha \nonumber \\
&=& i e^{\text{T}} \gamma^2 \gamma^0 \gamma^\mu e \nonumber \\
&=& -\bar{e}^c\gamma_\mu e \nonumber \\
&=& 0.
\end{eqnarray}
Note also that $\gamma^5\sigma^{\mu\nu}=2i\varepsilon^{\mu\nu\alpha\beta}
\sigma_{\alpha\beta}$ implies that $\bar{e}^c\gamma^5\sigma_{\mu\nu}e$
also vanish identically.  In Ref.\cite{Pas:2000vn}, these operators
were incorrectly included in their super-formula \footnote{However, they
neglected them in their final analysis because they worked in the
s-wave approximation.}.

Other color singlet operators that could potentially contribute to
$\nbb$-decay are
\begin{eqnarray}
\label{LOquarkop6}
{\cal{O}}_{6+}^{++} &=& (\bar{q}_{\text{L}}^a \tau^+ 
q_{\text{R},a}) (\bar{q}_{\text{R}}^b \tau^+
q_{\text{L},b})= \frac{1}{6} {\cal{O}}_{1+}^{++} , 
\\
\label{LOquarkop7}
{\cal{O}}_{7\pm}^{++} &=&  (\bar{q}_{\text{R}}^a\tau^+ \sigma^{\mu\nu}
q_{\text{L},a})(\bar{q}_{\text{R}}^b\tau^+\sigma_{\mu\nu} q_{\text{L},b} )
 \pm (\bar{q}_{\text{L}}^a \tau^+ \sigma^{\mu\nu}
q_{\text{R},a})(\bar{q}_{\text{L}}^b\tau^+\sigma_{\mu\nu}
q_{\text{R},b})=  \frac{12}{7} {\cal{O}}_{2\pm}^{++},
\\
\label{LOquarkop8}
{\cal{O}}_{8+}^{++} &=& (\bar{q}_{\text{L}}^a \tau^+ \sigma^{\mu\nu}
q_{\text{R},a})( \bar{q}_{\text{R}}^b\tau^+\sigma_{\mu\nu} q_{\text{L},b})=0, 
\\
\label{LOquarkop9}
{\cal{O}}_{9\pm}^{++,\mu} &=& (\bar{q}_{\text{L}}^a \tau^+ \sigma^{\mu\nu}
q_{\text{R},a} +  \bar{q}_{\text{R}}^a\tau^+ \sigma^{\mu\nu}
q_{\text{L},a} ) (\bar{q}_{\text{L}}^b\tau^+\gamma_\nu
q_{\text{L},b} \pm \bar{q}_{\text{R}}^b\tau^+\gamma_\nu q_{\text{R},b})=
\frac{-i}{1\pm 8} {\cal{O}}_{4\pm  }^{++,\mu},
\\
\label{LOquarkop10}
{\cal{O}}_{10\pm}^{++,\mu} &=& (\bar{q}_{\text{L}}^a \tau^+ \sigma^{\mu\nu}
q_{\text{R},a} -  \bar{q}_{\text{R}}^a\tau^+ \sigma^{\mu\nu}
q_{\text{L},a} ) (\bar{q}_{\text{L}}^b\tau^+\gamma_\nu
q_{\text{L},b} \mp \bar{q}_{\text{R}}^b\tau^+\gamma_\nu q_{\text{R},b})=
\frac{-i}{1\mp 8} {\cal{O}}_{5\pm}^{++,\mu},
\end{eqnarray}
where the latin indices denote color and terms that involve the
product of color octet currents are ignored (see below).
Using Fierz transformations and the following formula,
\begin{eqnarray}\label{colorrel}
\delta_{ab}\delta_{cd}=\frac{1}{3}\delta_{ad}\delta_{cb}
+\frac{1}{2}\sum\limits_{i=1}^{8}\lambda_{ad}^i\lambda_{cb}^i~,
\end{eqnarray}
it is easy to prove 
Eqs.~(\ref{LOquarkop6})-(\ref{LOquarkop10}).  Note that the second
term on the right-hand side of \Eq{colorrel} represents the product of
two color octet currents.  This term does
not contribute since the asymptotic states are colorless and a
completeness relation involving only hadronic states can be 
inserted between the currents.  We therefore neglect this
contribution.

Even though two
Fierz-related operators can arise due to different short-distance
dynamics, they are physically indistinguishable.  Note that in
Ref.\cite{Pas:2000vn}, these indistinguishable operators were
included as separate operators.

\section{Na{\"\i}ve dimensional analysis scaling rule}\label{nda}

To determine the scaling rules of the various fields appearing in the chiral
Lagrangian, start with the relation between the axial current and the
pion decay constant\cite{Donoghue:dd},
\begin{eqnarray}
\langle 0| A^{a,\mu} |\pi^b(p)\rangle = i\delta^{ab}f_\pi p^\mu,
\end{eqnarray}
which implies that $\pi$ is normally normalized by $f_\pi$.
Recalling that chiral perturbation theory is an expansion in powers of
$p/\lh$, we scale pion derivatives by $\lh$ noting that pion loop
corrections will involve factors of 
$p^2/(4\pi f_\pi)^2$; this suggests that $\lh\approx 4\pi f_\pi$.

Since the action is dimensionless, we also have from the kinetic energy
term of the pion field
\begin{eqnarray}
\int \text{d}^4\!x \partial^\mu\vec{\pi}\cdot\partial_\mu\vec{\pi}
= \int \text{d}^4\!x (\lh f_\pi)^2
\frac{\partial^\mu}{\lh}\frac{\vec{\pi}}{f_\pi} \cdot
\frac{\partial_\mu}{\lh}\frac{\vec{\pi}}{f_\pi}~.
\end{eqnarray}
This shows that we can associate with $\text{d}^4\!x$ the scale $(\lh
f_\pi)^2$.  This is the origin of the last factor of \Eq{georgi}.
From the parity-conserving pion nucleon coupling, we have
\begin{eqnarray}
\int \text{d}^4\!x
\frac{g_{\text{A}}}{f_\pi}\bar{N}\gamma^5\notder\pi N
= \int \text{d}^4\!x (\lh f_\pi)^2
\frac{g_{\text{A}}}{\lh f_\pi^2}\bar{N}
\gamma^5\frac{\notder}{\lh}\frac{\pi}{f_\pi} N.
\end{eqnarray}
This shows that we can associate the scale $\lh f_\pi^2$ with
$\bar{N}N$.

Next, we note that since the axial current at the quark level is given by
$\bar{q}\gamma^5\gamma^\mu q$ while a contribution to the axial
current at the hadronic level is $\bar{N}\gamma^5\gamma^\mu N$,  we can
also associate with $\bar{q}q$ the scale $\lh f_\pi^2$.  For a $\nbb$-decay
quark-lepton operator, this implies
\begin{eqnarray}
\frac{\gf^2}{\lbb}\int \text{d}^4\!x
(\bar{q}\Gamma q)  (\bar{q}\Gamma^\prime q)
(\bar{e}\Gamma^{\prime \prime} e^c)
= \frac{\gf^2f_\pi^2}{\lbb}\int \text{d}^4\!x (\lh f_\pi)^2
\frac{\bar{q}\Gamma q }{\lh f_\pi^2} \frac{\bar{q}\Gamma^\prime q }{\lh
  f_\pi^2}
\bar{e}\Gamma^{\prime \prime} e^c.
\end{eqnarray}
Therefore, we can associate the scale $\gf^2f_\pi^2/\lbb$ with the
lepton bilinears.  This explains the origin of the scaling rule in
\Eq{georgi}.

\section{NLO nuclear operators}\label{appnloops}

Here we present the results for Figs.~\ref{longrangediag}b and
\ref{longrangediag}c.  The Lagrangian \Eq{nnpieeefflag} gives

\begin{eqnarray}\label{diagbplusc}
\text{(b)+(c)} &=& 4i\frac{g_{\text{A}}M\lh}{\sqrt{2}}
\bar{u}_{e1} \gamma^2\gamma^0(\zeta_1+ \zeta_2\gamma^5)
\bar{u}_{e2}^{\text{T}}
\nonumber\\
& &~~~~~~~\times\left[\frac{(\bar{u}_{p3} u_{n1})
(\bar{u}_{p4}\gamma^5 u_{n2})}{(q_2^2 - m_\pi^2 + i\epsilon)} +
\frac{(\bar{u}_{p3}\gamma^5 u_{n1})
(\bar{u}_{p4}u_{n2})}{(q_1^2 - m_\pi^2 + i\epsilon)}\right]
\nonumber\\& &
+~4i\frac{g_{\text{A}}M}{\sqrt{2}f_\pi}
\bar{u}_{e1}\gamma_\mu\gamma^2\gamma^0
\gamma^5\bar{u}_{e2}^{\text{T}} \times
\nonumber\\
& &\hspace{1cm} \left[
\frac{ (\bar{u}_{p4}\gamma^5 u_{n2})}{(q_2^2 - m_\pi^2 + i\epsilon)}
 \bar{u}_{p3}\left(\zeta_3 + \zeta_4\gamma^5 \right)\gamma^\mu u_{n1}
~+ \right.
 \nonumber\\
& &\left. \hspace{1.5cm}~~\frac{ (\bar{u}_{p3}\gamma^5 u_{n1})}{(q_1^2 -
m_\pi^2 + i\epsilon)}
\bar{u}_{p4}\left(\zeta_3  + \zeta_4\gamma^5\right) \gamma^\mu u_{n2}
\right].
\end{eqnarray}
After taking the non-relativistic limit and performing a Fourier
transform we obtain:
\begin{eqnarray}\label{approxbplusc}
\text{F.T.(\ref{diagbplusc})}
&\simeq& \frac{1}{2\pi}
\frac{m_\pi }{\sqrt{2}g_{\text{A}}\lh}
\frac{g_{\text{A}}^2\lh^2}{\Lambda_{\beta\beta}^5}
\delta(\vec{x}_1-\vec{x}_3) \delta(\vec{x}_2-\vec{x}_4)
\frac{\text{e}^{-x}}{\rho}(1+\frac{1}{x})
\times
\nonumber \\
& &~~\left\{
~\bar{u}_{e1} \gamma^2\gamma^0(\zeta_1+ \zeta_2\gamma^5)
\bar{u}_{e2}^{\text{T}} ~~
(\delta_{24} \chi_3^\dagger
\vec{\sigma}\cdot\hat{\rho} \chi_1
-\delta_{13} \chi_4^\dagger
\vec{\sigma}\cdot\hat{\rho} \chi_2)
\right.
\nonumber \\
& &
~~~~~+\bar{u}_{e1} \gamma_\mu\gamma^2\gamma^0
\gamma^5\bar{u}_{e2}^{\text{T}}
\nonumber\\
& &~~~~~~~~~~
\times \left[-\chi_3^\dagger(\zeta_3 \delta^{\mu 0}-\zeta_4
\sigma^i \delta^{\mu i})\chi_1~~ \chi_4^\dagger
\vec{\sigma}\cdot\hat{\rho} \chi_2 \right.
\nonumber \\
& &
\left.\left.
~~~~~~~~~~~~~~+ \chi_4^\dagger(\zeta_3 \delta^{\mu0}-\zeta_4
\sigma^i \delta^{\mu i})\chi_2~~ \chi_3^\dagger
\vec{\sigma}\cdot\hat{\rho} \chi_1 \right] ~\right\}.
\end{eqnarray}
One can check explicitly that this nuclear operator is parity-odd and
does not contribute to the $0^+\to0^+$ nuclear transitions.  Note also
the extra factor of $m_\pi/\lh$ relative to the LO contribution of
\Eq{approxa} which is consistent with the power counting of
\Eq{pcounting}.

\end{appendix}



\end{document}